\documentclass[letterpaper]{JHEP3}

\usepackage{amssymb}
\usepackage{amssymb}
\usepackage{amsmath}
\usepackage{graphicx}
\usepackage{bm} 
\usepackage{amssymb}
\usepackage{amstext}
\usepackage{tensor}
\usepackage{epsf,epsfig}
\usepackage{amsfonts} 
\usepackage{graphicx}
 \usepackage{bm} 

\usepackage{amsmath}
\usepackage{mathrsfs}
 
\usepackage{amsmath,amssymb}

\newcommand{\insertplot}[5]{\begin{figure}
 \hfill\hbox to 0.05in{\vbox to #5in{\vfill
 \inputplot{#1}{#4}{#5}}\hfill}
 \hfill\vspace{-.1in}
 \caption{#2}\label{#3}
 \end{figure}}
 \newcommand{\inputplot}[3]{
 \special{ps: plotfile #1}
}
\newcounter{fig}   \newcommand{\lbfig}[1]{\refstepcounter{fig}

\label{#1} }

\newcommand{\beq}{\begin{equation}}
\newcommand{\eeq}{\end{equation}}
\newcommand{\beqs}{\begin{eqnarray}}
\newcommand{\eeqs}{\end{eqnarray}}

\newcommand{\be}{\begin{equation}}
\newcommand{\ee}{\end{equation}}
\newcommand{\bea}{\begin{eqnarray}}
\newcommand{\eea}{\end{eqnarray}}

\newcommand{\identity}{{\upright\rlap{1}\kern 2.0pt 1}}

\newcommand{\re}[1]{(\ref{#1})}

\numberwithin{equation}{section}

\abstract{ 
We consider the $\mathrm{O}(3)$ non-linear sigma-model,
composed of three real scalar fields with a standard kinetic term and with a symmetry breaking potential
in four spacetime dimensions.
We show that this simple, geometrically motivated model,
admits both self-gravitating, asymptotically flat, non-topological solitons and hairy black holes,
when minimally coupled to Einstein's gravity,
 $without$ the
need to introduce higher order kinetic terms in the scalar fields action.
Both spherically symmetric and spinning, axially symmetric solutions are studied.
The solutions are obtained under a
ansatz with oscillation (in the static case) or rotation (in the spinning case) in the internal space.  Thus, there is \textit{symmetry non-inheritance}: the matter sector is not invariant under the individual spacetime isometries. For the hairy black holes, which are necessarily spinning, the internal rotation (\textit{isorotation}) must be synchronous with the rotational angular velocity of the event horizon.
We explore the domain of existence of the solutions
and some of their physical properties, that resemble closely
those of  (mini) boson stars and Kerr black holes with synchronised scalar hair in Einstein-(massive, complex)-Klein-Gordon theory.
}

\preprint{ }

\title{    
Gravitating solitons and black holes with synchronised hair  in the
four dimensional
$\mathrm{O}(3)$ sigma-model
}

\author{{\large }
{\large C. Herdeiro}$^{\dagger}$,
{\large I. Perapechka}$^{\ddagger}$,
{\large E. Radu}$^{\ast\Diamond}$,
and {\large Ya. Shnir}$^{\star\bullet}$
\\ \\
$^{\ddagger}${\small
Centro de Astrof\'\i sica e Gravita\c c\~ao - CENTRA, Departamento de F\'\i sica,
Instituto Superior T\'ecnico - IST, Universidade de Lisboa - UL, Avenida
Rovisco Pais 1, 1049-001, Portugal
}
\\
$^{\ddagger}${\small Department of Theoretical Physics and Astrophysics, Belarusian State
University,
Nezavisimosti Avenue 4, Minsk 220004, Belarus}
\\
$^{\ast}${\small School of Theoretical Physics, Dublin Institute for Advanced Studies,
 10 Burlington Road, Dublin 4, Ireland}
\\
$^{\Diamond}${\small CIDMA, Universidade de Aveiro,
Campus de Santiago, 3810-183 Aveiro, Portugal}
\\
$^{\star}${\small 
BLTP, JINR, Dubna 141980, Moscow Region, Russia}
\\
$^{\bullet}${\small 
Department of Theoretical Physics, Tomsk State Pedagogical University, Russia}
}

\begin{document}

\section{Introduction}

Non-linear sigma models are scalar field theories wherein the scalar fields
take values on a certain \textit{target manifold}, described as a non-linear function of the fields.
They were first introduced long ago~\cite{GellMann:1960np} in the context of a field theoretical description of pion decay.
A conceptually simple, but physically and mathematically rich particular example is the $\mathrm{O}(3)$ sigma model, which
 has the very simple Lagrangian density
\be
\mathcal{L}=\frac{1}{2}\partial_\mu \phi^\alpha \,  \partial^\mu \phi_\alpha \ ,
\label{o3sigma}
\ee
where the trio of scalar fields, $\phi^\alpha=(\phi^1,\phi^2,\phi^3)$ parameterises a 2-sphere: $\phi^\alpha\phi_\alpha=1$.
Thus, on Euclidean 3-space, $\mathbb{R}^3$, the set of $\phi^\alpha$ define a map $\mathbb{R}^3\longrightarrow S^2$; moreover, the fields must (spatially) asymptote to constant values, which may be chosen, say, as  $\phi^\alpha_\infty=(0,0,1)$.
The identification of $\mathbb{R}^3$'s spatial infinity as a single point effectively replaces $\mathbb{R}^3$ by its one point compactification, $S^3$. Then, the scalar fields become a map
\be
\phi^\alpha: S^3\longrightarrow S^2 \ ,
\ee
which is naturally characterised by the \textit{Hopf index}, the third homotopy group of $S^2$, $\pi_3(S^2)=\mathbb{Z}$.

As it turns out, this very simple, geometrically appealing field theory has no finite energy, localized
solutions on  $\mathbb{R}^3$, as shown by using a standard Derrick-type scaling argument~\cite{Derrick:1964ww}, $cf.$ Section~\ref{nogosec}.
Such \textit{solitonic} configurations are, nonetheless, found  by augmenting the Lagrangian~\eqref{o3sigma}
with a Skyrme-type term~\cite{Skyrme:1961vq} which is quartic in the derivatives of the scalar fields, as first realized in~\cite{Fadeev1,Faddeev:1976pg}.
This results in the Faddeev-Skyrme model, which has been extensively studied over the last 30 years,
its solutions carrying a \textit{topological} charge given by the (integer) Hopf index
and  being usually called in the
literature {\it Hopf solitons} or {\it Hopfions}. Hopfions have found a variety of
applications in different branches of science, including not only physics~\cite{Kauffman,Ackerman,Sutcliffe-2017,Sutcliffe-2018},
but also chemistry \cite{Knots} and biology \cite{Sumners}.

\bigskip

With the exception of adding higher derivative terms, no other mechanism to endow the $\mathrm{O}(3)$-non-linear sigma model (\ref{o3sigma})
with finite mass-energy solutions is known. The addition of a (positive) potential term $V$, in particular, is insufficient for
the existence of solitonic solutions, as shown, again by a scaling argument, $cf.$ Section~\ref{nogosec}; and  adding rotation
to the model -- which results in isospinning Hopfions with some angular frequency $w$ in the Faddeev-Skyrme model --
still requires the higher order kinetic term for the existence localised, finite energy solutions~
\cite{Harland:2013uk,Battye:2013xf}.

\bigskip

It is interesting to contrast the situation we have just described with the picture found in a different class of non-linear scalar field theories wherein \textit{non-topological} solitons in flat space exist - \textit{$Q$-balls}~\cite{Friedberg:1976me,Coleman:1985ki}.
These solitons emerge in the complex-Klein-Gordon field theory
with a mass term and a self-interacting potential, which, in the simplest cases, contains a quartic and sextic term (besides the quadratic mass term). The emergence of these solutions, evading Derrick type no-soliton arguments~\cite{Derrick:1964ww},
is intrinsically linked to the existence of harmonic oscillations
in field space that do not carry through into the energy-momentum dynamics.
This is an early example of \textit{symmetry non-inheritance}:
the matter fields do not share the full symmetry observed at the level of the
spacetime energy-momentum distribution \cite{Smolic:2015txa}.

\bigskip

It is well known that $Q$-balls self-gravitate,
when their complex-Klein-Gordon field theory is minimally coupled to Einstein's gravity. When the
non-linearities of gravity are introduced, moreover, the non-linearities of the field theory become~\textit{optional}.
Self-gravitating scalar solitons in the Einstein-complex-Klein-Gordon model only require a mass term.
These self-gravitating, asymptotically flat, everywhere regular lumps of scalar
 field energy are called \textit{boson stars},
and were originally obtained without scalar self-interactions~\cite{Kaup:1968zz,Ruffini:1969qy}.
Such boson stars, obtained in a model solely with a mass term are referred to as
\textit{mini} boson stars~\cite{Schunck:2003kk}. Boson stars containing scalar self-interactions,
of quartic type, were first considered in~\cite{Colpi:1986ye};  boson stars with the simplest $Q$-ball type potential
were discussed in~\cite{Kleihaus:2005me,Kleihaus:2007vk}.
This latter example yields non-trivial flat spacetime configurations ($Q$-balls)
as the gravitational coupling is switched off. The former examples of boson stars, trivialise in that limit.
In all cases, the models allows both static, spherically symmetric, and stationary, spinning, axially
symmetric boson star solutions - see $e.g.$~\cite{Yoshida:1997qf,Schunck:1996he,Grandclement:2014msa,Herdeiro:2016gxs}.

We remark that boson stars, like $Q$-balls, exhibit symmetry non-inheritance. In the case of
spinning boson stars, the complex scalar field is, strictly speaking neither stationary nor axially symmetric.
In fact, there is an explicit temporal and azimuthal dependence of the complex scalar field, endowing it with
a \textit{phase rotation}; but the corresponding energy momentum tensor is stationary and axially symmetric,
and therefore the required compatibility with a spacetime geometry preserved by these isometries is fulfilled.

\bigskip

In a recent development, it has been shown that each of these \textit{spinning}, asymptotically flat
boson star models belong to a larger family of solutions of \textit{hairy black holes (BHs)}~\cite{Herdeiro:2014goa,Herdeiro:2015gia}. That is, it is always possible to place a BH horizon in equilibrium with a rotating boson star, as long as the phase rotation of the boson star is synchronised with the angular velocity of the BH horizon. Thus, these are BHs with \textit{synchronised hair}. Taking the limit where the horizon goes to zero size within this family of solutions,  the hairy BHs reduce to spinning boson stars; in the limit where the scalar field trivialises, they reduce to the vacuum Kerr BH of general relativity~\cite{Kerr:1963ud}, for the particular BH parameters that can support \textit{test field stationary scalar clouds}~\cite{Hod:2012px,Hod:2013zza,Herdeiro:2014goa,Hod:2014baa,Benone:2014ssa,Hod:2016lgi}. These hairy BHs violate Wheeler's \textit{dynamical} no hair conjecture~\cite{Ruffini:1971bza} as they can form dynamically~\cite{East:2017ovw,Herdeiro:2017phl} and  be sufficiently long lived~\cite{Degollado:2018ypf}. Moreover, in the space of solutions, they can have phenomenological features quite distinct from Kerr, including different shadow shapes and topologies~\cite{Cunha:2015yba,Vincent:2016sjq} and a distinct $X$-ray spectroscopy~\cite{Ni:2016rhz,Franchini:2016yvq}. Also, they present some distinct geometrical features, such as new shapes of ergo-regions~\cite{Herdeiro:2014jaa} and an interior geometry distinct from (eternal) Kerr~\cite{Brihaye:2016vkv}. The way in which these BHs circumvent well known no hair theorems~\cite{Bekenstein:1972ny} is precisely related to the symmetry non-inheritance they share with the spinning boson stars they reduce to in the vanishing horizon limit - see $e.g.$~\cite{Smolic:2015txa,Herdeiro:2015waa}.

\bigskip

The case of mini-boson stars shows that the coupling of a field theory wherein no solitonic solutions exist to gravity makes such finite energy lump solutions possible. This suggest that, similarly to the case of the massive, free, complex Klein-Gordon field theory,
minimally coupling the $\mathrm{O}(3)$ non-linear sigma model~\eqref{o3sigma} to Einstein's gravity may yield solitonic finite energy solutions,   $without$ a quartic (or higher order) term  in the action. Moreover, since one is now in the realm of gravity, BH solutions, and in particular hairy BHs, should exist.

 \bigskip

The main purpose of this work is to show that, indeed, the coupling  of
the $\mathrm{O}(3)$ non-linear sigma model (\ref{o3sigma})
 to Einstein's gravity results in families of  solitonic and hairy BH solutions
which closely resemble the pattern found in the non-self interacting, massive, complex scalar field case.
Again, the existence of both the   solitons and the hairy BHs relies on a symmetry non-inheritance mechanism.
The scalar fields are neither stationary nor axisymmetric, but possess a rotation in the field space - \textit{isorotation} (or oscillations, in the static case).
When the angular velocity of this isorotation matches that of the black hole horizon, hairy BHs are possible. In all cases, the oscillations in field space imply, for the existence of bound states,  a potential must be added to the matter Lagrangian~\eqref{o3sigma}, $cf.$ eq.~\eqref{lagNLS} below, which plays the role of a mass term.
 Both the solitonic solutions and hairy BHs trivialise in the flat spacetime limit,
as they are topologically trivial, for the (non-generic) ansatz we consider.

\bigskip

This paper is organised as follows. In Section~\ref{section1} we introduce the model, including the field equations, conserved Noether current and Noether charge. Scaling type arguments are then used to establish the absence of solitonic solutions in flat spacetimes. Then, the spacetime and scalar fields ansatz, as well as the boundary conditions to be used in finding the numerical solutions are presented, together with some relevant physical quantities to be analysed. In Section~\ref{section2} we discuss our numerical results. Firstly we discuss the domain of existence of the spherical and spinning solitons; next, then the domain of existence of the hairy BHs is analysed. A discussion of some physical properties, including the matter distribution around the BHs, the BH temperature, the types of ergo-regions observed and the shape of the horizon follows. We also consider the variation of the solutions with the gravitational coupling and illustrate the discrete set of families of solutions that arise in these models, labelled by an integer - the azimuthal winding number. Finally, in Section~\ref{section3} we present some conclusions and further remarks.

\section{The Model}
\label{section1}

\subsection{Field equations and current}

We consider the non-linear ${O}(3)$ sigma model minimally coupled to Einstein's gravity in 3+1 dimensions. The model's action is
\be
\label{lag}
\mathcal{S} = \int{\sqrt{-g}\left(\frac{R}{16 \pi G}-\mathcal{L}_\mathrm{m}\right) d^4 x}\ ,
\ee
where $R$ is the scalar curvature, $g$  is the determinant of the metric tensor, $ G$ is Newton's constant,
and $\mathcal{L}_\mathrm{m}$ is the matter field Lagrangian:
\be
\label{lagNLS}
\mathcal{L}_\mathrm{m} = \frac{\lambda_1}{2}\left(\partial_\mu \phi^\alpha\right)^2+\lambda_0
\left(1-\phi^3\right)\, ,
\ee
where the trio of real scalar fields $\phi^\alpha$, $\alpha = 1, 2,3$, is restricted to the surface of the unit sphere:
\be
(\phi^1)^2+(\phi^2)^2+(\phi^3)^2=1 \ .
\label{constraint}
\ee
In order to yield asymptotically flat solutions, the boundary condition at spatial infinity are $\phi^\alpha \rightarrow \phi^\alpha_\infty=(0,0,1)$. The target space of the $\mathrm{O}(3)$ sigma model is therefore  $S^2$.
Also, $\lambda_0$ and $\lambda_1$ are
(dimensionful)
 input parameters,
their ratio fixing the mass of the fields' excitations,
$\mu^2=\lambda_0/\lambda_1$.
Observe also that by appropriately rescaling the coordinates $x^\mu$ and $G$ one can effectively set  $\lambda_0=\lambda_1 = 1$,
leaving only one non-trivial parameter, $\alpha$,
where we define $\alpha^2\equiv 4\pi G \lambda_1$.

Variation of \re{lag} with respect to metric yields the Einstein equations:
\be
\label{Einstein}
R_{\mu\nu}-\frac{1}{2}R g_{\mu\nu}=2\alpha^2 T_{\mu\nu}\ ,
\ee
where the stress-energy tensor  is
$$
T_{\mu\nu}=\partial_\mu \phi^\alpha \partial_\nu \phi^\alpha  - \mathcal{L}_\textrm{m}g_{\mu\nu} \ .
$$
Variation of \re{lag} with respect to scalar field itself leads to the following  field equations:
\be
\label{scaleq}
\partial_\mu \partial^\mu\phi^\alpha +  \phi^\alpha_\infty=0\ ,
\ee
which are solved together with the constraint equation (\ref{constraint}).

Observe that the potential term in \eqref{lagNLS} breaks the $\mathrm{O}(3)$ symmetry of the model to the $\textrm{SO}(2)$ subgroup.
Associated to the latter,
a Noether current exists, given by
\be
\label{Noether}
j_\mu = -\phi^1 \partial_\mu \phi^2 + \phi^2 \partial_\mu \phi^1\ .
\ee

\subsection{Absence of flat spacetime solitons}
\label{nogosec}
Let us first establish that the field theory~\eqref{lagNLS} in flat spacetime admits no finite energy,
localised solutions, even in the presence of a generic, positive potential $V[\phi^3]$,  using a scaling
argument. We consider the spacetime action on Minkowski space, $\mathbb{M}^{1,3}$
\begin{eqnarray}
\mathcal{S}_{\mathbb{M}^{1,3}}=\int dt \mathcal{I}_{\mathbb{R}^3} \ ,\qquad  \mathcal{I}_{\mathbb{R}^3}=\int_{R^3}d^3 x
\left(
\frac{1}{2}\partial_i \phi^a \,  \partial^i \phi^a
+V[\phi^3]
\right)
\equiv
I_2+I_0 \ ,
\label{En1}
\end{eqnarray}
where $i,k$ are spatial indices and  $I_2,I_0$  are positive quantities.
Following Derrick \cite{Derrick:1964ww}, let us assume there is a non-trivial solution, and
we consider the scale transformation thereof,
$x^i\to \Lambda x^i$
(with $\Lambda$ an arbitrary constant), defining a 1-parameter family of configurations.  This scaling implies
$I_2  \to \Lambda I_2$
and
$I_0 \to \Lambda^3 I_0$.
Then, requiring the original configuration is a solution yields
$(dI(\Lambda)/d\Lambda)|_{\Lambda =1}=0$,
which implies the virial identity
\begin{eqnarray}
 I_2+3 I_0=0 \ .
\end{eqnarray}
Since both terms are positive definite, this implies they both must vanish, and the hypothetical solution must be trivial.

One may inquire if the presence of a harmonic time dependence, which is key to the existence of $Q$-balls, could change
the above result, and yield flat spacetime solitons.
As shown by Ward \cite{Ward:2003un} (see also \cite{Mareike}),  in three spatial dimensions
the answer is still negative (at least for the usual form of the potential) and can be proven as follows.
Without any loss of generality, one takes an stationary $\mathrm{O}(3)$ ansatz with factorized time dependence:
$\phi^1+i \phi^2=\psi (x^k)e^{-i\omega t}$ (with $\psi$ a complex function in general) and $\phi^3(x^k)$.
Then, \eqref{En1} becomes
\begin{eqnarray}
\label{nv}
 \mathcal{I}_{\mathbb{R}^3}=\int_{R^3}d^3 x
\left(
\frac{1}{2}
(
|\nabla \psi|^2+(\nabla \phi^3)^2
-\omega^2 |\psi|^2
)
 +V[\phi^3]
\right) \ .
\end{eqnarray}
Recall that, differently from the static case, the presence of a potential
(with the associated mass term)
is a necessary requirement for possible bound states, when $\omega \neq 0$.
Restricting to the potential considered above
\begin{eqnarray}
\label{pot1}
V[\phi^3]=\mu^2 (1-\phi^3)  \ ,
\end{eqnarray}
(with $ \mu$ the mass parameter)
and following the same Derrick prescription as in the static case, one arrives at the virial identity
\begin{eqnarray}
\label{nv2}
 \int_{R^3}d^3 x
\left\{
\partial_i \phi^a \,  \partial^i \phi^a
 +6 (1-\phi^3)\left[\mu^2-\frac{\omega^2}{2}(1+\phi^3)\right]
\right\}=0\ .
\end{eqnarray}
Observing that $w<\mu$, which follows from a a linearised analysis
of the solutions in the far field and the fact that $|\phi^3|<1$, one concludes that the integrand in (\ref{nv2})
is strictly positive, which, again, rules out the existence of solitonic configurations in $d=3$ spatial dimensions.\footnote{On the
other hand, there are stable spinning soliton solutions in the pure $d=2$ $\mathrm{O}(3)$ sigma model with a polynomial potential
\cite{Ward:2003un,Mareike}.}
Moreover,
this conclusion is independent of the precise spatial symmetries of the configurations. It should be remarked, however, that this scaling argument does not exclude the existence of solitonic configurations for more
complicated potentials than (\ref{pot1}); in fact we predict such solutions to exist. In the following, in order to find solitonic solutions of
the $\mathrm{O}(3)$ sigma model,~\eqref{lagNLS} with the potential~(\ref{pot1}), we shall consider its coupling to gravity, $i.e.$ the action~\eqref{lag}.

\subsection{Ansatz}
We seek stationary, axially-symmetric solutions of \re{Einstein}-\re{scaleq}, describing spinning,
asymptotically flat solitons or ``hairy" BHs. Using coordinates adapted to the two commuting
Killing vectors $\xi=\partial_t$ and $\eta=\partial_\varphi$, where  $t$ and $\varphi$ are
the time and azimuthal coordinates, the metric can be written in Lewis-Papapetrou form:
\be
\label{metrans}
ds^2=-F_0 dt^2 +F_1\left(dr^2+r^2 d\theta^2\right)+ r^2\sin^2 \theta F_2  \left(d\varphi-\frac{W}{r} dt\right)^2,
\ee
where the four metric functions, $F_0, F_1, F_2$ and $W$, depend on $r$ and $\theta$ only.

The ${O}(3)$ scalar field ansatz compatible with stationarity
and axial symmetry \textit{needs not} to be $t,\varphi$ independent.
The solutions in this work are found within an ansatz\footnote{We remark that the most general  $\mathrm{O}(3)$ scalar field ansatz
compatible with the isometries of (\ref{metrans})  contains an extra function for the $(\phi^1,\phi^2)$-sector.}
inspired by previous work done for an $\mathrm{O}(4)$ Skyrme model
\cite{Battye:2005nx,Ioannidou:2006nn,Battye:2014qva,Shnir:2016qbz,Perapechka:2017bsb,Herdeiro:2018daq},
with:
\be
\label{scalans}
\phi^\alpha =
\left[
\sin f\cos\left(m\varphi-\omega t\right),~
\sin f\sin\left(m\varphi-\omega t\right),~
\cos f
\right]\ ,
\ee
where the profile function $f$ depends on coordinates $r$ and $\theta$ only,
$\omega$ is the
spinning frequency of field - there are \textit{isorotations} - and $m\in\mathbb{Z}$ is the azimuthal
harmonic index. Observe that the trigonometric parameterisation \eqref{scalans} identically obeys
the sigma model constraint \eqref{constraint}.

As a special case we shall also be considering static, spherically symmetric self-gravitating solitons without an event horizon;\footnote{Similar solutions of the  $\mathrm{O}(4)$ sigma model were considered in \cite{Verbin}.} spherically symmetric hairy BHs turn out not to exist.
In this limit, the line element (\ref{metrans}) has  $W=0$ while $F_1,F_2,F_0$ depends only on $r$ (with $F_1=F_2$).
The scalar field ansatz is of the form~\eqref{scalans} with $m=0$ and $f$ solely a radial function.
For the remaining of this Section,
however, we will focus on the generic stationary case and the metric ansatz~\eqref{metrans}.

We assume the existence of a rotating, topologically spherical event horizon, located at a constant value of radial variable
$r=r_h>0$. The horizon null generator is the  helicoidal Killing vector field
\be
\label{Killingrh}
\chi=\xi +\Omega_H \eta\, \ ,
\ee
where the
horizon angular velocity,  $\Omega_H$, is fixed by the value of the metric function $W$ on the horizon:
$$
\Omega_H =  -\frac{g_{\phi t}}{g_{tt}}\biggl.\biggr|_{r=r_h} = W\biggl.\biggr|_{r=r_h} \ .
$$
The  rotating horizon allows the existence of stationary scalar clouds,
supported by the synchronisation condition \cite{Herdeiro:2014pka,Herdeiro:2014goa,Herdeiro:2014ima}
\be
\label{synchron}
\omega = m \Omega_H \ ,
\ee
between the event horizon angular velocity, $\Omega_H$ and the angular phase velocity $\omega/m$ of the scalar field. This condition implies that there is no scalar field through the horizon \cite{Herdeiro:2014pka,Herdeiro:2014goa,Herdeiro:2014ima}.

The $\mathrm{SO}(2)$ unbroken symmetry yields the conserved
Noether current $j^\mu$, $cf.$ \re{Noether}. Thus, there is an associated Noether charge
\be
\label{Noetcharge}
Q=\int{\sqrt{-g}j^0 d^3 x}
=2\pi\int_{r_h}^\infty dr \int_0^\pi
 d\theta{\frac{F_1\sqrt{F_2}\sin^2 f }{\sqrt{F_0}}
r^2\sin\theta \left(\omega -\frac{m W}{r}\right)}\ ,
\ee
which is the counterpart of the Noether charge of boson stars, associated with the global phase rotations of the
complex scalar field~\cite{Friedberg:1986tp,Kleihaus:2005me,Herdeiro:2014pka}.

\subsection{Boundary conditions and relevant physical quantities}

In the generic stationary case it is convenient to make use of the following exponential parametrization for the metric fields
\be
\label{subst}
F_0=\frac{\left(1-\frac{r_h}{r}\right)^2}{\left(1+\frac{r_h}{r}\right)^2}e^{f_0},
\quad F_1=\left(1+\frac{r_h}{r}\right)^4 e^{f_1}
\quad F_2=\left(1+\frac{r_h}{r}\right)^4 e^{f_2}\, .
\ee
Then a power series expansion near the horizon yields the following regularity requirements for the profile function
$f(r)$ and the metric functions $f_i$
\be
\label{bchor}
\partial_r f\bigl.\bigr|_{r=r_h}=\partial_r f_0\bigl.\bigr|_{r=r_h}=\partial_r f_1\bigl.\bigr|_{r=r_h}
=\partial_r f_2\bigl.\bigr|_{r=r_h} = 0\, ,
\ee
These boundary conditions supplement the synchronization condition \re{synchron} imposed on the metric function
$W$.

The requirement of asymptotic flatness  at spatial infinity yields another set of boundary conditions:
\be
\label{bcinf}
f\bigl.\bigr|_{r\to \infty}=f_0\bigl.\bigr|_{r\to \infty}
=f_1\bigl.\bigr|_{r\to \infty}=f_2\bigl.\bigr|_{r\to \infty}
=W\bigl.\bigr|_{r\to \infty}=0\, .
\ee
Axial symmetry and regularity impose the following boundary conditions
on the symmetry axis at $\theta=0,\pi$,
\be
\label{bcpole}
f\bigl.\bigr|_{\theta = 0,\pi} =
\partial_\theta f_0\bigl.\bigr|_{\theta = 0,\pi} =
\partial_\theta f_1\bigl.\bigr|_{\theta = 0,\pi} =
\partial_\theta f_2\bigl.\bigr|_{\theta = 0,\pi} =
\partial_\theta W\bigl.\bigr|_{\theta = 0,\pi}=0 \ .
\ee
We also require solutions to be $\mathbb{Z}_2$-symmetric under reflections with respect to the equatorial plane. Thus, it is enough to consider the range of values
of the angular variable $\theta \in [0,\pi/2]$. The corresponding boundary conditions on the equatorial plane are:
\be
\label{bcaxis}
\partial_\theta f\bigl.\bigr|_{\theta =
\frac{\pi}{2}} = \partial_\theta f_0\bigl.\bigr|_{\theta =
\frac{\pi}{2}} = \partial_\theta f_1\bigl.\bigr|_{\theta=\frac{\pi}{2}}
= \partial_\theta f_2\bigl.\bigr|_{\theta=\frac{\pi}{2}}
= \partial_\theta W\bigl.\bigr|_{\theta=\frac{\pi}{2}}=0 \ .
\ee

Furthermore,  the absence of a conical singularity at the symmetry axis
requires that the deficit angle should vanish,
$\delta=2\pi\left(1 - \lim\limits_{\theta\to 0} \frac{F_2}{F_1}\right) = 0$.
Hence any  physically consistent solution should satisfy the constrain
$F_2\bigl.\bigr|_{\theta = 0}=F_1\bigl.\bigr|_{\theta = 0}$.
In our numerical scheme we explicitly imposed this condition on the symmetry axis.

Asymptotic expansions of the metric function at the horizon and at spacial infinity yields a number of physical
observables.
The total ADM mass $M$ and the angular momentum $J$ of the spinning hairy BH
can be read off from asymptotic subleading behaviour of metric functions as $r\to \infty$:
\be
\label{ADM}
g_{tt}=-1 + \frac{2 MG}{r}+\mathcal{O}\left(\frac{1}{r^2}\right), \qquad
g_{\varphi t}=-\frac{G J}{r}\sin^2 \theta+\mathcal{O}\left(\frac{1}{r^2}\right) \ .
\ee

The ADM charges  can be decomposed as sum of two contributions, one from the event horizon and one from the bulk scalar
hair: $M = M_H+M_\Phi$ and $J = J_H+J_\Phi$, respectively. These contributions
can be evaluated separately using Komar integrals\footnote{All numerical work herein uses 
$\lambda_1=\lambda_0=1$ and $ \alpha^2=4 \pi G$.}
\be
\label{Komar}
\begin{split}
M_H&=-\frac{1}{8\pi G}\oint_\mathrm{S}{dS_{\mu\nu} \nabla^\mu \xi^\nu}\ ,
\quad J_H=\frac{1}{16\pi G}\oint_\mathrm{S}{dS_{\mu\nu} \nabla^\mu \eta^\nu}\ ,\\
M_\Phi&=- \int_{V}{dS_{\mu} \left(2 T^\mu_\nu \xi^\nu - T\xi^\mu\right)}\ ,
\quad J_\Phi= \int_\mathrm{V}{dS_{\mu} \left(T^\mu_\nu \eta^\nu - \frac{1}{2} T\eta^\mu\right)}\ ,
\end{split}
\ee
where $S$ is the horizon 2-sphere and $V$ denotes an asymptotically flat spacelike hypersurface bounded by the horizon.

We remark that similarly to the angular momentum of the stationary rotating boson stars \cite{Kleihaus:2005me,Kleihaus:2007vk},
there is a quantisation relation for the angular momentum of the scalar field,
$J_\Phi=m Q$, where $Q$ is the  Noether charge  $Q$ \re{Noetcharge} and $m$ is the winding number of the scalar field.

The relevant horizon quantities include the Hawking
temperature $T_H$, which is  proportional to the surface gravity
$\kappa^2=-\frac{1}{2}\nabla_\mu\chi_\nu\nabla^\mu\chi^\nu$,
as
\be
\label{horQ}
T_H=\frac{\kappa}{2\pi} = \frac{1}{16\pi r_h}\exp\left[\frac{1}{2}\left(f_0 - f_1\right)\bigl.\bigr|_{r = r_h}\right]\, .
\ee
Here $\chi$ is the horizon null generator \re{Killingrh}. Another quantity of interest is the horizon area which is given by
\be
\label{horQ-A}
A_H = 32\pi r_h^2 \int_0^\pi d\theta \sin\theta\exp\left[\frac{1}{2}\left(f_1 + f_2\right)\bigl.\bigr|_{r = r_h}\right] \, .
\ee 
One can easily check that the horizon quantities are related through the Smarr relation
\be
\label{Smarr}
M = 2T_H S + 2\Omega_H J_H + M_\Phi \ ,
\ee
where 
$S=\frac{1}{4 G}A_H$ is the BH entropy  and $M_\Phi$ is the scalar field energy outside the
event horizon, \re{Komar}.
Another relation between the physical quantities of the hairy BH is provided by the first law of thermodynamics
$$
dM = T_H dS + \Omega_H d J \ .
$$

\section{Numerical results}
\label{section2}

For the spinning solutions, we solve the boundary value problem for nonlinear partial differential equations
\re{Einstein}-\re{scaleq} with boundary conditions \re{bchor}-\re{bcaxis} using a
fourth-order finite differences scheme. The system of equations is discretized on a grid
with $201\times 101$ points. To simplify our calculation in the near horizon area, we introduce
a new radial coordinate $x=\frac{r-r_h}{r+c}$, which maps the semi-infinite region $[0,\infty )$ onto the unit interval
$[0,1]$. Here $c$ is an arbitrary constant which is used to adjust the contraction of the  grid.
The emerging system of nonlinear algebraic equations is solved using a trust-region Newton method \cite{Lin}.
The underlying linear system is solved with the Intel MKL PARDISO sparse direct solver \cite{pardiso}.
Errors are of order of $10^{-4}$. 
Most calculation are performed using the CESDSOL\footnote{Complex Equations -- Simple
Domain partial differential equations SOLver is a C++ package
being developed by one of us (I.P.).
} library. Some
of the solutions were also constructed by using FIDISOL/CADSOL package \cite{schoen}
which also uses the Newton-Raphson method.
But let us start with the spherical solutions for which the numerical strategy is simpler.

\subsection{Spherical solitons - domain of existence}

 The solitonic solutions are found by fixing $r_h=0$.
Thus there are only two continuous input  parameters $(\alpha,\omega)$ and one discrete one, $m$.
 For $m=0$ we find spherically symmetric solitons.
Although they can be studied in isotropic coordinates
($i.e.$ the static limit of (\ref{metrans})),
we found it convenient to employ Schwarzschild-like coordinates,
with a line element
\begin{eqnarray}
ds^2=-N(r)\sigma^2(r) dt^2+\frac{dr^2}{N(r)}+r^2(d\theta^2+\sin^2 \theta d\phi^2)\ , \qquad
N(r)\equiv 1-\frac{2m(r)}{r}\ .
\label{sphericalansatz}
\end{eqnarray}
Then, for
 the scalar fields ansatz~\eqref{scalans}, the problem reduces to solving a second order equation for the scalar amplitude $f(r)$
and two first order equations for the metric functions $m$ and $\sigma$
\begin{eqnarray}
&&
(r^2 N\sigma f')'=r^2\sigma \left(\mu^2-\frac{\omega^2}{N\sigma^2}\cos f\right)\sin f\ ,
\\
&&
m'=\frac{1}{2}\alpha^2 r^2\left[
N f'^2+\frac{\omega^2 \sin^2 f}{N\sigma^2}+4\mu^2 \sin^2\left(\frac{f}{2}\right)
\right] \ ,
\\
&&
\sigma'= \alpha^2 r \sigma
\left(
f'^2+\frac{\omega^2 \sin^2 f}{N^2\sigma^2}
\right) \ .
\end{eqnarray}
The equations for a spherically symmetric (mini) boson star are recovered  to lowest order
in the limit of a small $f$.
Differently from that case, however,  the sigma model constraint prevents us from absorbing the coupling constant $\alpha$
in the expression of the scalars.

Close to the origin, the approximate form of the functions read
\begin{eqnarray}
\label{origin}
&&
f(r)=b+\frac{1}{6}\left[\mu^2-\frac{\omega^2}{\sigma^2}\cos b\right]\sin b r^2+\mathcal{O}(r^4) \ ,
\\
\nonumber
&&
m(r)=\frac{2}{3}\alpha^2\left[\mu^2+\frac{\omega^2}{\sigma^2}\cos^2\left(\frac{b}{2}\right)\right]\sin^2\left(\frac{b}{2}\right)+\mathcal{O}(r^5)\ ,
\\ \nonumber
&&
\sigma(r)=\sigma_0+\frac{\alpha^2 \omega^2\sin^2b}{2\sigma_0}r^2+\mathcal{O}(r^4) \ ,
\end{eqnarray}
where $b$ and $\sigma_0$
two free parameters.
The leading order terms in the  large-$r$ solutions for the various functions  are
\begin{eqnarray}
\label{infinity}
&&
 f(r)=\frac{f_0e^{-\sqrt{\mu^2-\omega^2}r}}{r}+\dots, \qquad m(r)=M-\frac{\alpha^2 f_0^2 e^{-2\sqrt{\mu^2-\omega^2}r}\mu^2}{2 \sqrt{\mu^2-\omega^2}}+\dots,~~
\\
\nonumber
&&
\sigma(r)=1-\frac{\alpha^2 f_0^2 }{r} \sqrt{\mu^2-\omega^2}  e^{-2\sqrt{\mu^2-\omega^2}r} +\dots,
\end{eqnarray}
where $M$ is the ADM mass and $f_0$ another parameter, both fixed by the numerics. In this case the  Noether charge takes the form
\begin{eqnarray}
Q= 4 \pi \omega\int_0^{\infty}dr~r^2 \frac{\sin^2F}{N\sigma} \ .
\end{eqnarray}

The numerical construction of the solutions is straightforward.
We use a standard
Runge-Kutta ordinary differential equation solver and evaluate the initial conditions at
$r =10^{-6}$ for global
tolerance
$r =10^{-14}$
 adjusting for fixed shooting parameters
$f(0)$, $\sigma(0)$
and integrating towards $r\to  \infty$.
The accuracy of the solutions was also
monitored by computing the virial identity
\begin{eqnarray}
\label{virial-grav}
\int_0^{\infty}dr~r^2 \sigma
\left[
f'^2+12 \mu^2\sin^2\left(\frac{f}{2}\right)
\right]
=
\omega^2\int_0^{\infty}dr
\frac{r^2(4N-1)\sin^2 f}{N^2\sigma}\ ,
\end{eqnarray}
which shows that, as with mini boson stars, the coupling to gravity is crucial for the existence of these solutions; indeed  (\ref{virial-grav}) cannot be satisfied for a flat metric, as for $N=\sigma=1$ it (\ref{virial-grav}) becomes
\be
\int_0^{\infty}dr~r^2
\left\{
f'^2+12 \sin^2\left(\frac{f}{2}\right)\left[\mu^2-\omega^2\cos^2\left(\frac{f}{2}\right)\right]
\right\}
= 0 \ ,
\ee
which cannot be satisfied since $\omega<\mu$.

\begin{figure}[ht]
\hbox to\linewidth{\hss%
    \resizebox{8cm}{6cm}{\includegraphics{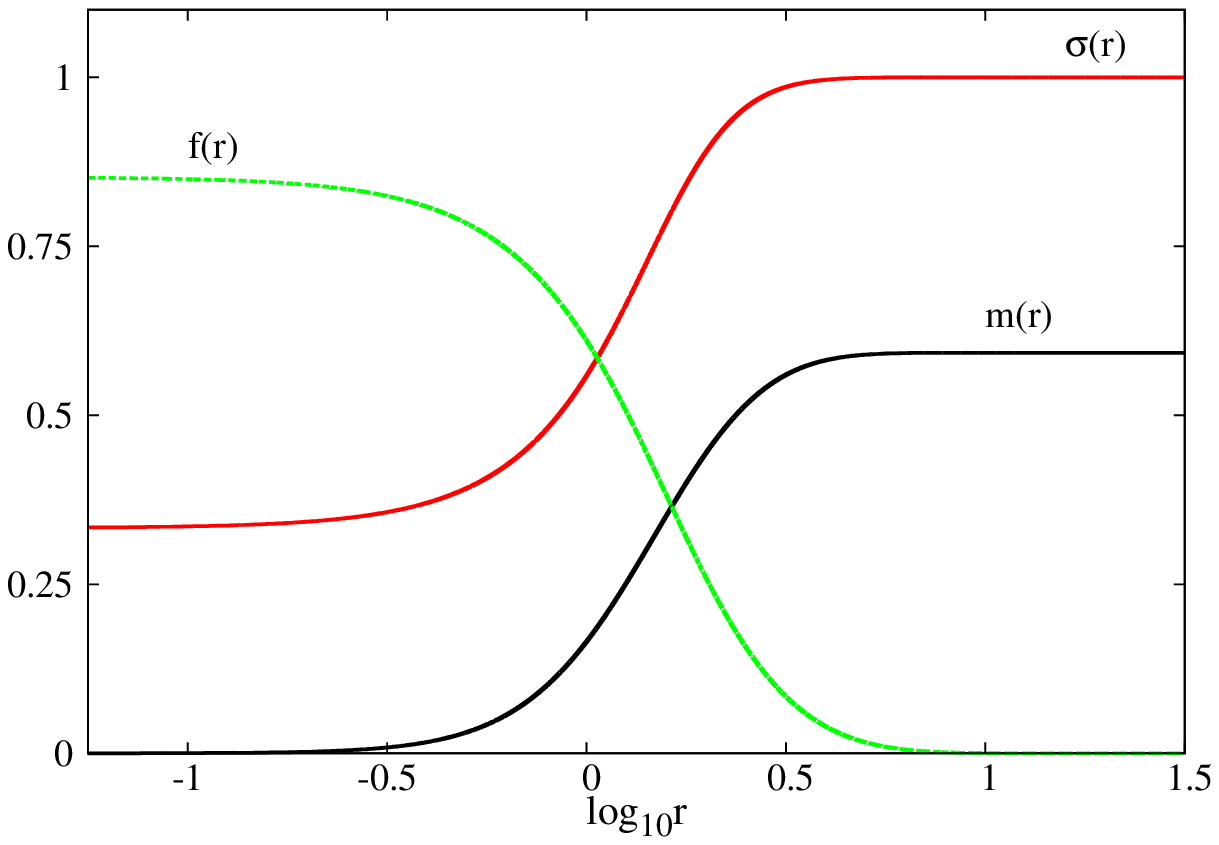}}
\hspace{10mm}%
        \resizebox{8cm}{6cm}{\includegraphics{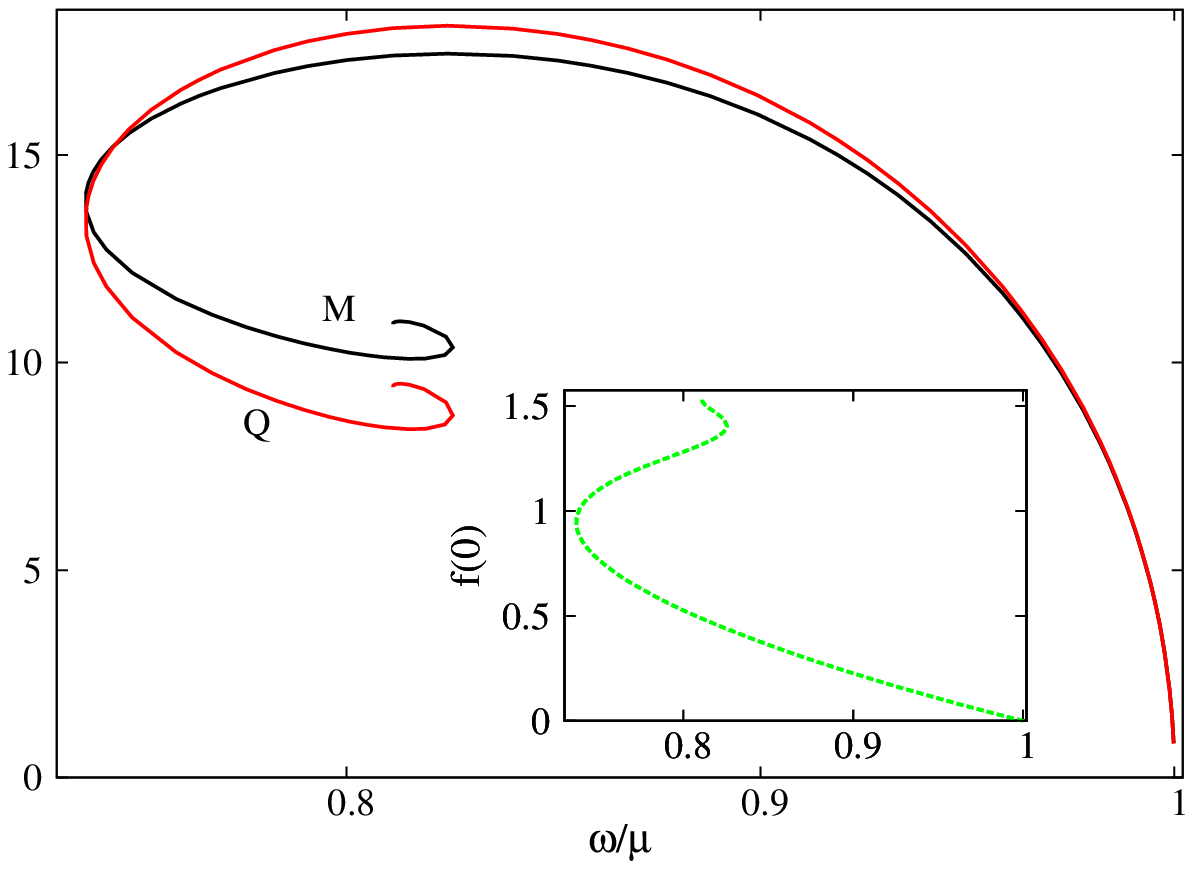}}
\hss}
\caption{\small
Spherically symmetric gravitating $\mathrm{O}(3)$ solitons. (Left panel)
The radial profile functions of an illustrative solution, which has $w/\mu=0.74$.
(Right panel) The  mass and Noether charge as function of $\omega$. For both panels, $ \alpha=1/\sqrt{2}$.
}
   \lbfig{spherical}
 \end{figure}

As usual with boson stars, there are fundamental states, for which the scalar fields profile has no nodes and excited states, labelled by the number of such nodes. Indeed, for  given $\omega,\alpha$, the solutions form a discrete
set, labelled by the number of nodes, $n$, of the function $f$. Here and also in the spinning case we will always focus on fundamental modes. A typical profile is shown in Fig.~\ref{spherical} (left panel),
for a nodeless solution.

The domain of existence of such solitons, in an ADM mass $vs.$ frequency diagram is shown in Fig.~\ref{spherical} (right panel),
for the illustrative value
$\alpha=1/\sqrt{2}$
and $n=0$. The plot also shows the Noether charge and the value of the scalar fields at the origin.
This domain of existence  corresponds to a spiral, a typical pattern for boson star solutions and other gravitating solitons. The solitons exist for some restricted set of values of the angular frequency
$\omega\in\left[\omega_\mathrm{min} \simeq 0.737,\omega_\mathrm{max}=1\right]$.
 The upper critical value $\omega_\mathrm{max}=1$ corresponds to the
particular choice of the potential of the model \re{lagNLS}. This can be confirmed by introducing a complex scalar field $\Psi$ as $\Psi=\phi^1+i\phi^2$. Then the matter Lagrangian~\eqref{lagNLS} with the rescaling done before and taking into account the sigma model constraint~\eqref{constraint} becomes, to first order in $\Psi$:
\be
\label{linearfield}
\mathcal{L}_{\rm m}\simeq \frac{1}{2}\partial_\mu\Psi \partial^\mu \Psi^*+\frac{\mu^2}{2}|\Psi|^2 \ ,
\ee
where the effective mass used in numerics 
is $\mu=1$. Since $\omega^2\leqslant \mu^2$ is a bound state condition, the maximal frequency is $\omega_{\rm max}=1$. In this limit the spinning solitons smoothly approach perturbations around Minkowski spacetime, with the ADM mass $M$ tending to zero. Indeed, linearizing the scalar field equation \re{scaleq} one can see that the profile function decays asymptotically as $f\sim e^{-\sqrt{1-\omega^2}r}$, becoming delocalized as $\omega\to 1$. This is usually referred to as the \textit{Newtonian limit}.

Starting from the Newtonian limit in Fig.~\ref{spherical} (right panel),
 in which limit the function $f(r)$ becomes very close to zero
and the solution trivialises, at some intermediate frequency $\omega_{\rm Mmax}$, a maximal mass is attained.
Similarly to the mini boson stars case, there is also a minimal
frequency, $\omega_{\rm min}$, below which no solutions are found. The solutions in between the Newtonian limit and $\omega_\mathrm{min}$ compose the \textit{first (forward) branch}. At $\omega_\mathrm{min}$ the curve backbends and a \textit{second (backward) branch} is found. A multi-branch structure ensues, with a third (forward) branch visible in  Fig.~\ref{spherical} (right panel), providing an overall inspiral-type pattern for the line of solutions.
Likely, the spiral  approaches, at the  centre, a critical singular solution
with $f(0)=\pi/2$. We remark that, as a rule of thumb, as one advances along the spiral one is moving towards the strong gravity region where the solitons become more compact along the forward branches and slightly less compact along the backwards branches - see, $e.g.$ the bottom panels of Fig. 1 in~\cite{Cunha:2017wao}. On the other hand, the central value of the scalar field increases monotonically along the spiral, $cf.$ inset in  Fig.~\ref{spherical} (right panel), as in other boson star models, $cf.$ inset in bottom panels of Fig. 1 in~\cite{Cunha:2017wao}.

As a final remark concerning the spherical case, following the approach in~\cite{Pena:1997cy},
one can prove these spherical solitons do not possess generalizations with an event horizon at their center:
$i.e.$ there are no spherically symmetric BHs with $\mathrm{O}(3)$-sigma model hair.

\subsection{Spinning solitons - domain of existence}

Taking $m\geqslant 1$,  we obtain spinning solitonic solutions. The dependency of these solutions on the angular frequency $\omega$ is qualitatively similar to that just described for the spherical solitons and also observed in spinning boson star models - see the outermost solid blue line in the left panel of Fig.~\ref{omeganew}. Again we focus on fundamental modes. A recent study of excited spinning boson stars and Kerr BHs with synchronised hair has been discussed in~\cite{Wang:2018xhw}.

\begin{figure}[hbt]
    \begin{center}
        \includegraphics[width=.48\textwidth, trim = 40 20 90 20, clip = true]{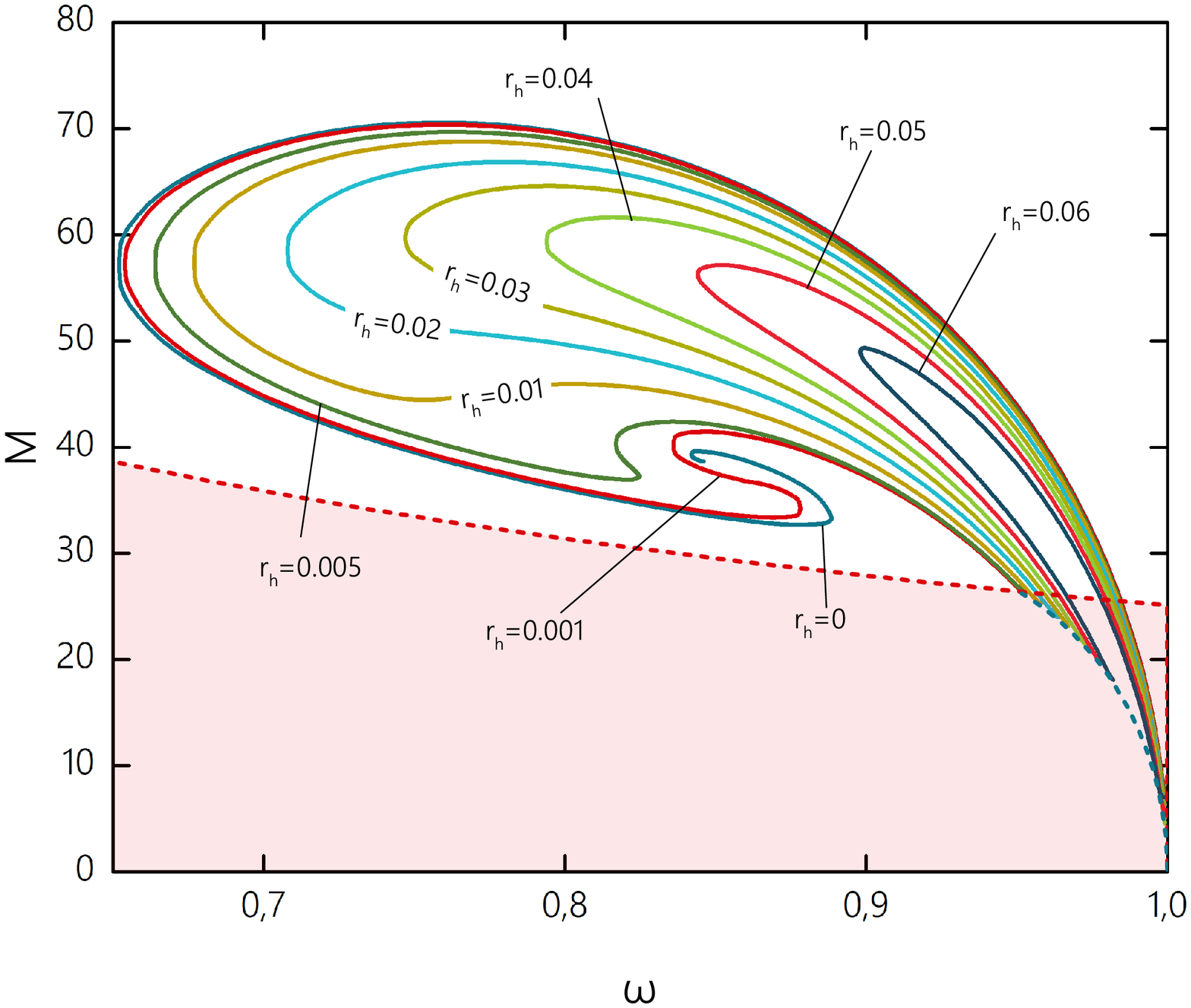}
        \includegraphics[width=.48\textwidth, trim = 40 20 90 20, clip = true]{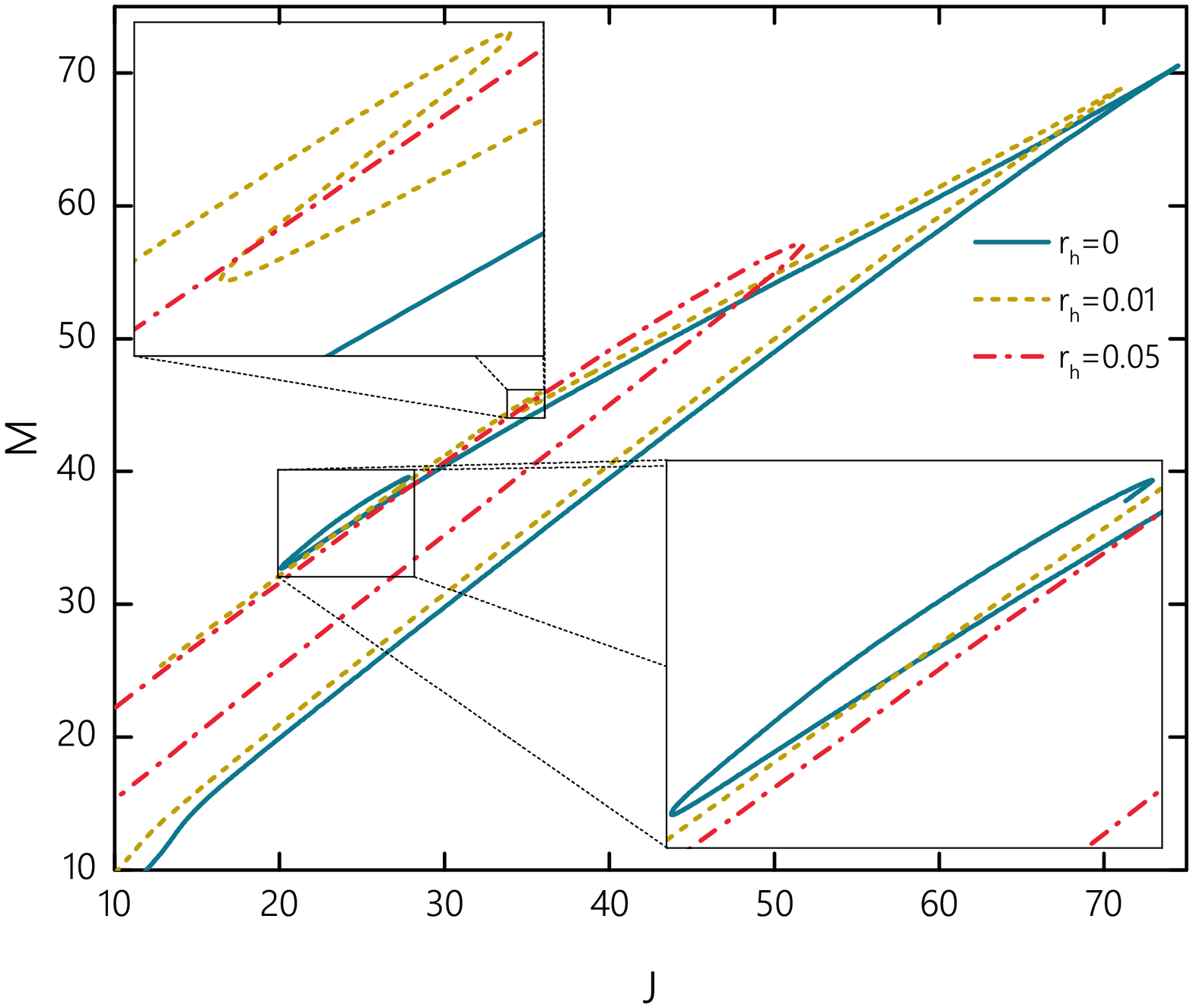}
    \end{center}
    \caption{\small
    ADM mass $M$ $vs.$  (left panel) the angular frequency $\omega$ or (right panel)  the angular momentum $J$. Here $m=1$ and $\alpha=0.5$, whereas $r_h$ varies. In the left panel, the shaded area corresponds to the domain of existence of vacuum Kerr BHs, the red dashed line to the extremal vacuum Kerr BHs, and the blue dashed line to the \textit{existence line}: the subset of vacuum Kerr BHs that can support test field, stationary scalar clouds of the linear, massive, Klein-Gordon equation, with zero nodes and angular quantum numbers $\ell=1=m$, of the spheroidal harmonics~\cite{Herdeiro:2014goa}.
    }
    \lbfig{omeganew}
\end{figure}

Starting from the Newtonian limit, we observe that, as the angular frequency decreases from $\omega_{\rm max}=1$, the ADM mass of all solutions increases  approaching its maximum at some value of frequency $\omega_{\rm Mmax}$. From that point onwards the mass decreases until the minimum frequency  $\omega_\mathrm{min}$ is reached. Using the same nomenclature as before, the solutions in between the Newtonian limit and $\omega_\mathrm{min}$ compose the first (forward) branch. At $\omega_\mathrm{min}$ the curve backbends into a second (backward) branch. A  third (forward) branch and a fourth (backward) branch are visible in Fig.~\ref{omeganew}, making up a spiral towards a limiting solution at the center of the spiral (challenging to obtain numerically).
This mimics closely what has been found for both rotating and non-rotating boson stars (see $e.g.$~\cite{Herdeiro:2015gia}) as well as for other gravitating solitons ($e.g.$ Proca stars~\cite{Brito:2015pxa}); the location and the shape of the spiral depend on the gravitational coupling strength $\alpha$ \cite{Friedberg:1986tp,Kleihaus:2005me,Kleihaus:2007vk}.

The blue solid curve in the right panel of~Fig.~\ref{omeganew} shows the ADM mass of the spinning solitons as a function of their total angular momentum. Again, the pattern follows closely that observed for spinning boson stars - see $e.g.$~\cite{Herdeiro:2015gia}. Along the first branch both mass and angular momentum increase until the maximal ADM mass is attained; then the trend inverts and both mass and angular momentum decrease until the minimum ADM mass attained along the second (backward) branch. Another inversion follows and so on, revealing a zig-zag type structure.

In summary,  the solitonic solutions of the $\mathrm{O}(3)$ sigma model mimic closely those of the more familiar mini boson stars.  Boson stars with a $Q$-ball type potential also have similar features, but admit a non-trivial flat space limit, where they become $Q$-ball solutions. The spinning solitons of the non-linear $\mathrm{O}(3)$ sigma model trivialise in the flat spacetime limit, as mini-boson stars do.

\subsection{Hairy black holes - domain of existence}

Let us now turn to the hairy black hole solutions we constructed numerically. The problem has four input parameters, three of them  continuous ($\alpha, r_h, \omega$) and one discrete $m$.
We have obtained more than 4000 solutions in order to study how the solitons and the
hairy BHs depend on each of these parameters. To simplify our analysis, we mainly consider the spinning
hairy BHs with winding number $m=1$, but below we shall also exhibit some solutions with $m\geqslant 1$.

The left panel of  Fig.~\ref{omeganew} exhibits the variation of the ADM mass of the  spinning hairy BHs,
at a fixed value of the gravitational coupling $\alpha=0.5$, versus the
angular frequency $\omega$, along some lines of constant $r_h\neq 0$. First observe that the hairy BHs smoothly connect with the spinning solitons. Concerning the structure of the $r_h=$ constant lines, part of the description provided above for the solitonic limit still applies: we again observe that, as the angular frequency decreases from $\omega_{\rm max}$, the mass of all solutions increases
approaching its maximum at some value of frequency $\omega_{\rm Mmax}$. This maximal mass decreases
as $r_h$ increases. The behaviour of the $r_h=$constant curves beyond $\omega_{\rm Mmax}$ depends on the value of $r_h$.

For very small values of $r_h$ the spiral type critical
behavior observed in the solitonic limit changes. There is still a multibranch structure; but instead of terminating at some central limiting solution, the plot shows that after three backbendings a fourth branch terminates at an upper critical value of the angular frequency $\omega_{\rm end}<1$, see Fig.~\ref{omeganew} (left panel).
In this limit, the scalar field trivializes and we recover a Kerr BH solution with the corresponding value of $r_h$ and  that can support  $\mathrm{O}(3)$ non-trivial configurations as \textit{test field solutions}. This subset of Kerr BHs (as one varies $r_h$) defines the \textit{existence line}, which, since the field becomes small, by virtue of~(\ref{linearfield}) coincides with that found in the Einstein-(complex, massive)-Klein-Gordon theory. We call this the~\textit{Kerr limit}. This behaviour is qualitatively similar to that observed for other models of Kerr BHs with synchronised~\cite{Herdeiro:2014goa,Herdeiro:2015gia}.

As $r_h$ increases, the multibranch structure gives place to a two-branch scenario, with the first (upper, forward) branch connected to the perturbative excitations at $\omega_{\rm max}=1$
and the second (lower, backward) branch ending on the
Kerr solution as $\omega \to \omega_{\rm end}$.  The minimum value of the angular frequency $\omega_{\rm min}$ is increasing as $r_h$ increases; at some point the
frequency $\omega_{\rm Mmax}$ which corresponds to the maximal value of the ADM mass along the constant $r_h$ line, becomes the minimal allowed frequency $\omega_{\rm min}$. Also, the maximum value of the frequency along the second branch $\omega_{\rm end}$ is slowly increasing and it approaches $\omega_{\rm max}=1$ as the loop shrinks to zero. Fig.~\ref{omeganew} (left panel) presents the domain of existence we have just described. A very similar behaviour was recently described for excited Kerr BHs with synchronised scalar hair~\cite{Wang:2018xhw}. In the right panel of  Fig.~\ref{omeganew} we see the aforementioned zig-zag type structure remains for sequences of hairy BHs with $r_h={\rm constant}\neq 0$.

\subsection{Some physical properties of the solutions}
Having analysed the domain of existence of the solitons and hairy  BHs found in the $\mathrm{O}(3)$ sigma model, which in particular shows their global quantities $M,J$, let us now consider some other basic physical properties of the solutions.

In Fig.~\ref{profile} (left panel) we plot the scalar field profile function for two illustrative hairy BHs, both with the same value of $r_h$ but one on the first (forward) branch and another on the second (backward) branch. The figure illustrates two features. Firstly, that the scalar field distribution is the typical toroidal pattern of spinning boson stars, with a clear maximum along the equator at some radial distance and a suppression of the profile function for larger latitudes (the field vanishes along the symmetry axis). Secondly, that going along the spiral (away from the Newtonian limit), the solutions become more compact and the scalar field attains larger values.

\begin{figure}[hbt]
    \begin{center}
        \includegraphics[width=.48\textwidth, trim = 40 20 90 20, clip = true]{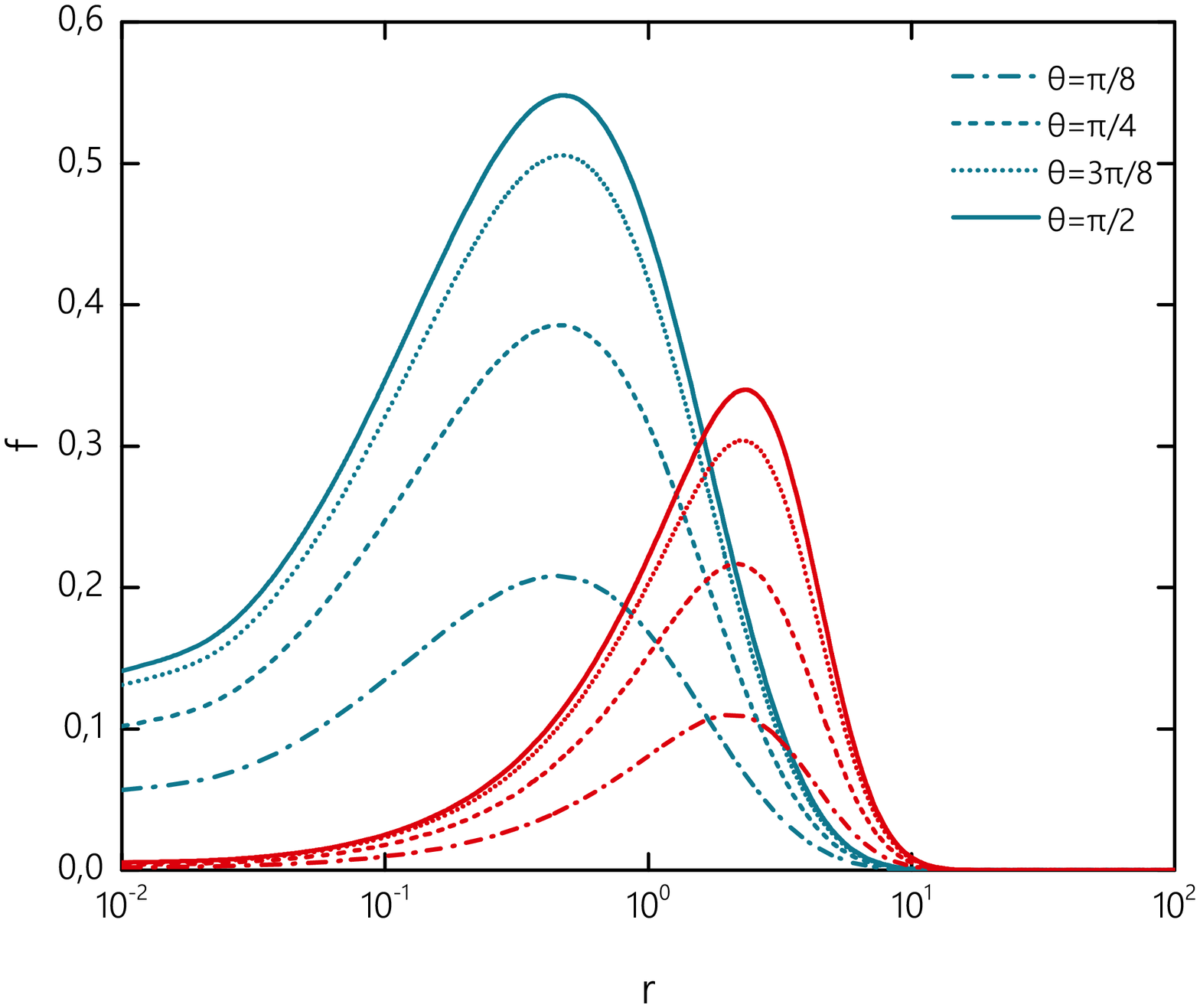}
           \includegraphics[width=.48\textwidth, trim = 40 20 90 20, clip = true]{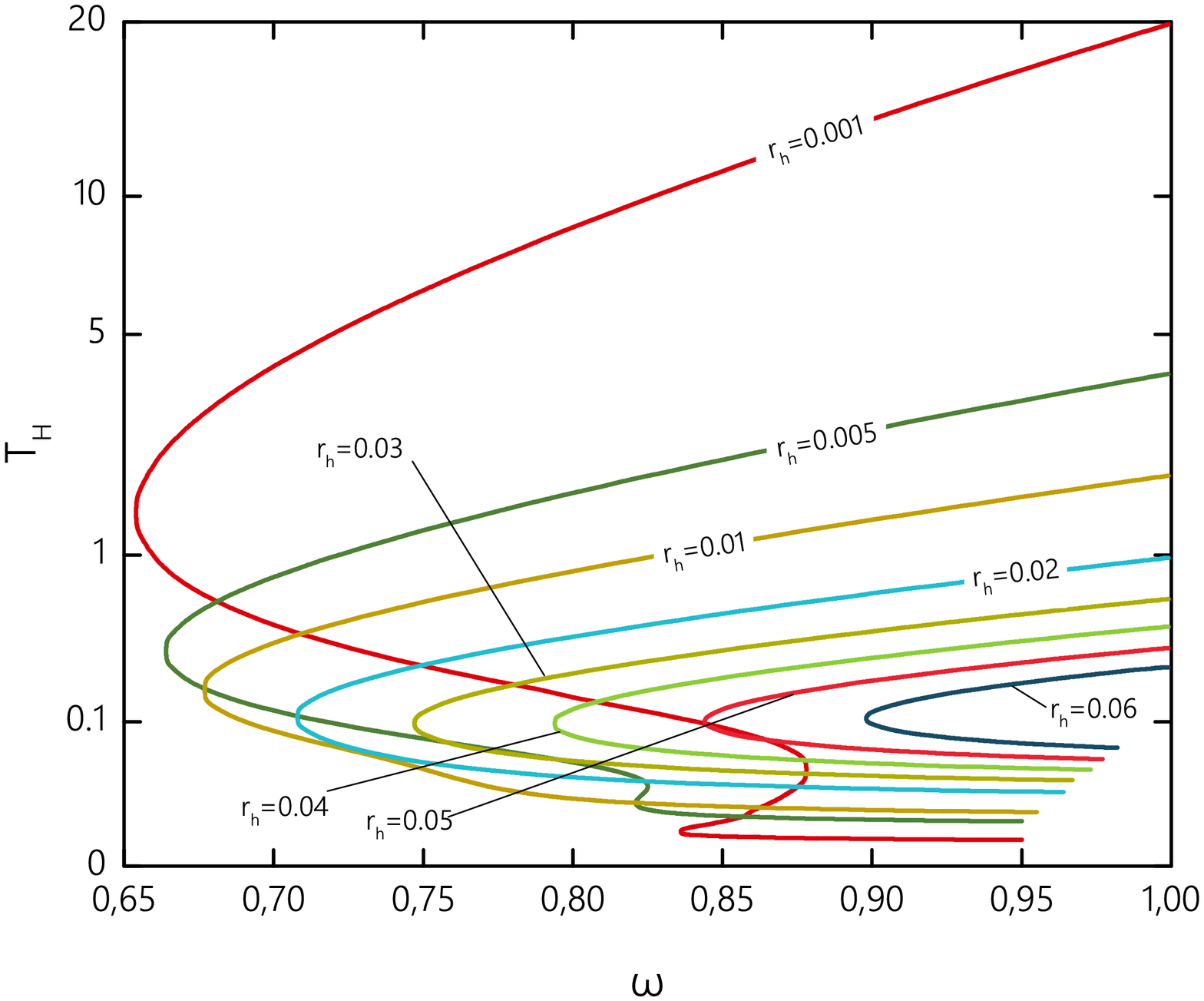}
    \end{center}
    \caption{\small
        (Left panel) Scalar field profile function $f(r,\theta)$ of two hairy BH solutions with $r_h=0.01,~w=0.8$,  one on the first forward branch (red curves) and another on the second backward branch (blue curves) branches, for several values of  $\theta=\textrm{const}$.         (Right panel) Hawking temperature $T_H$ of  hairy BHs, as function of the angular frequency for a set of values of the horizon radius $r_h$. All solutions have $\alpha=0.5$, $m=1$.}
    \lbfig{profile}
\end{figure}

In the right panel of Fig.~\ref{profile} we exhibit the Hawking temperature for the same solutions that were plotted in the domain of existence, $cf.$~Fig.~\ref{omeganew}.  One observes that, in the Kerr limit the solutions with the smallest $r_h$ have the smallest temperature. This is consistent with the distribution of the endpoint of the curves along the existence line, observed in the left panel of Fig.~\ref{omeganew}, where the smallest $r_h$ is the one closest to the extremal Kerr solutions ($cf.$ the $q=0$ line in Fig. 2 in~\cite{Herdeiro:2014goa}).  In the Newtonian limit, on the other hand, BHs become Schwarzschild like and $r_h$ is a good measure of their size. Then the largest BHs have the lowest temperature, as expected.

The toroidal structure of the matter distribution exhibited in the left panel of Fig.~\ref{profile} has another interesting manifestation. Similarly to what has been observed for spinning boson stars and Kerr BHs with synchronised hair~\cite{Herdeiro:2014jaa} toroidal \textit{ergo-regions} appear in the spinning solitons and hairy BHs of the $\mathrm{O}(3)$ model. These toroidal ergo-regions are delimited by an \textit{ergo-surface},   defined as the zero locus of the time-like Killing vector $\xi \cdot \xi =0$, or
\be
g_{tt}= -F_0 + \sin^2 \theta  F_2 W^2 = 0 \, .
\label{ergo-sf}
\ee
The analysis of the ergo-regions of the spinning boson in \cite{Herdeiro:2014jaa} showed that these solitons do not have an ergo-region in the vicinity of the Newtonian limit. Moving towards the strong gravity region, however, they develop an ergo-region still in the first forward branch, after the maximum ADM mass. The topology of the corresponding ergo-region is always an ergo-torus, $S^1\times S^1$. The hairy BHs -- that connect continuously to the spinning boson stars in this model --  can either have a Kerr-like ergo-region, delimited by a (topological) 2-sphere, $S^2$, or an \textit{ergo-Saturn}, with topology $S^2 \bigoplus (S^1\times S^1)$. The former (latter) hairy BHs can be thought as a superposition of a rotating horizon with a spinning boson star without (with) a toroidal ergo-region.

\begin{figure}[hbt]
    \begin{center}
        \includegraphics[width=.42\textwidth, trim = 0 10 0 10, clip = true]{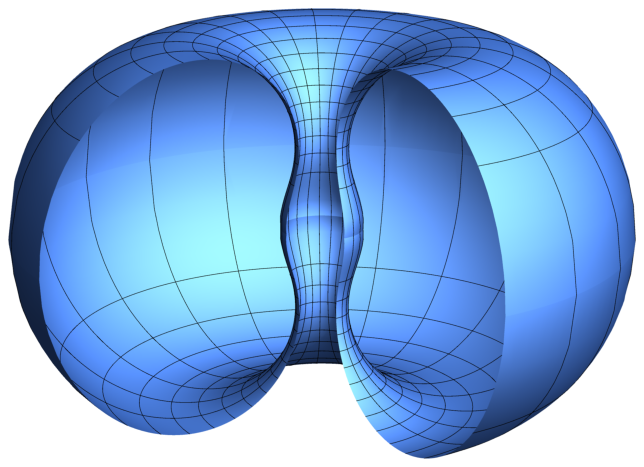}
        \includegraphics[width=.42\textwidth, trim = 0 10 0 10, clip = true]{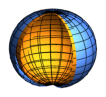}
        \includegraphics[width=.42\textwidth, trim = 0 10 0 10, clip = true]{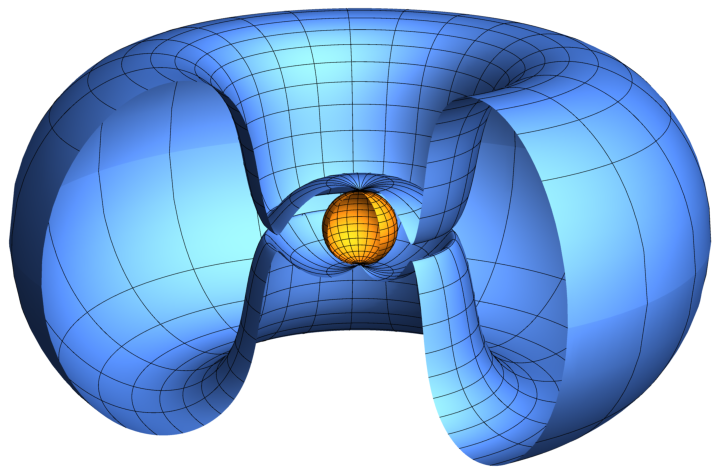}
        \includegraphics[width=.42\textwidth, trim = 0 10 0 10, clip = true]{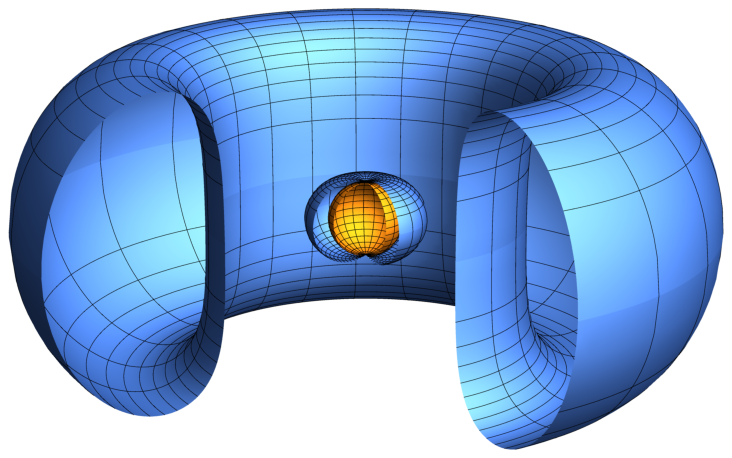}
    \end{center}
    \caption{\small Illustration of the ergo-surfaces of some selected solutions.  (Upper left panel) Spinning soliton along the second (backward) branch with $\omega = 0.8$; (upper right panel) hairy BH along the first (forward) branch with $r_h=0.01$ and $\omega = 0.8$; (lower left panel) hairy BH along second (backward) branch with $r_h=0.01$ and $\omega = 0.7$; (bottom right panel) hairy BH along the second (backward) branch with $r_h=0.01$ and $\omega = 0.685$. In the last three plots
    the event horizon is plotted as a yellow surface.}
    \lbfig{ergo}
\end{figure}

Using our numerical data we have found the solutions of  equation \re{ergo-sf} unveiling a qualitatively similar picture to that in~\cite{Herdeiro:2014jaa}. Firstly, we observe that on the forward branch, the axially-symmetric rotating boson stars with $r_h=0$
appear to possess no ergosurfaces. However, on the backward branch they develop an $S^1 \times S^1$ ergo-surface  (ergo-torus) -  Fig.~\ref{ergo}, left upper panel. This is a pattern also observed in other boson star models, with a $Q$-ball type potential \cite{Kleihaus:2005me,Kleihaus:2007vk}.

On the other hand, spinning hairy BH solutions of the $\mathrm{O}(3)$ model, both on the first forward branch and on most  of the
secondary branches possess a Kerr-like ergo-surface with  $S^2$ topology - Fig.~\ref{ergo}, right upper panel. At some point along the spiral, however, the spinning hairy BHs develop an ergo-torus in addition to the $S^2$ ergo-sphere, thus giving rise to an ergo-Saturn, just as in~\cite{Herdeiro:2014jaa} - Fig.~\ref{ergo}, bottom panels. A similar pattern is observed for the hairy BH solutions
of the Skyrme model recently analysed in~\cite{Herdeiro:2018daq}.

Moving on with our analysis, we now consider the variation of the solutions with the coupling $\alpha$. In Fig.~\ref{alpha} we exhibit the scaled mass $M\alpha^2$ of the hairy BH solutions at a fixed value of
the angular frequency $\omega$ versus the gravitational coupling constant $\alpha$.

\begin{figure}[hbt]
    \begin{center}
        \includegraphics[width=.48\textwidth, trim = 40 20 90 20, clip = true]{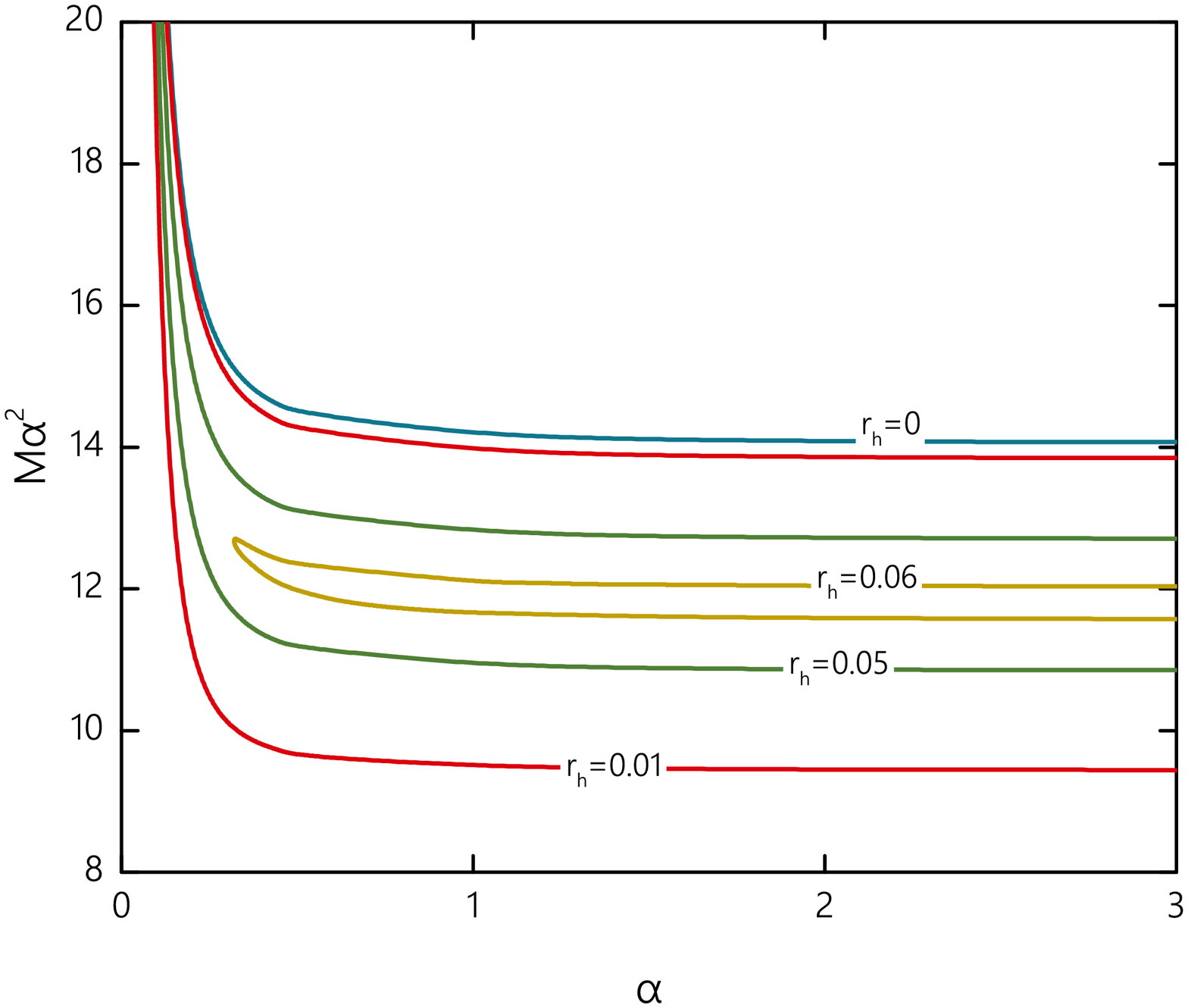}
        \includegraphics[width=.48\textwidth, trim = 40 20 90 20, clip = true]{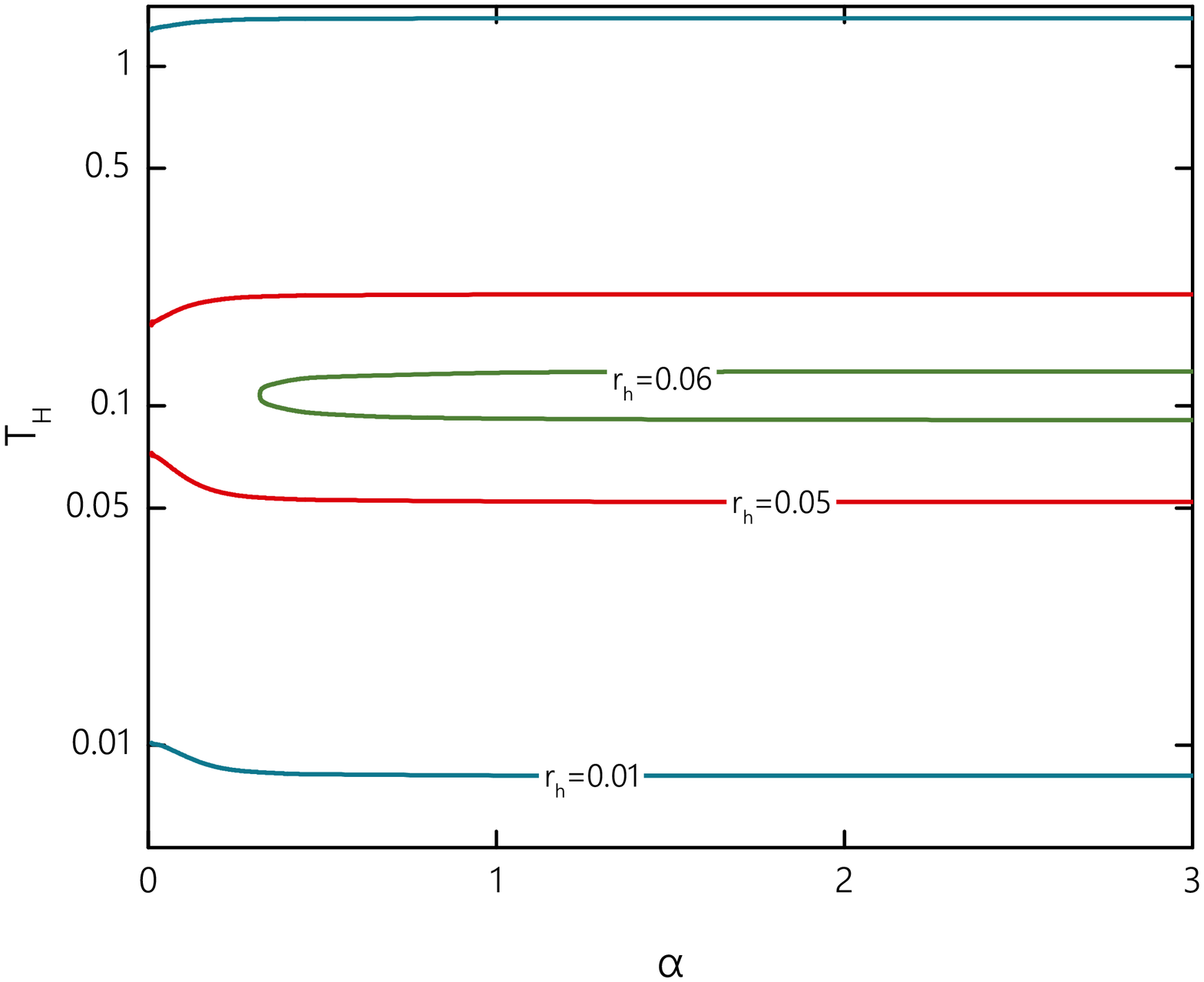}
    \end{center}
    \caption{\small
        Scaled ADM mass $M\alpha^2$ (left panel) and  Hawking temperature $T_H$ (right panel)
        $vs.$ the gravitational coupling constant $\alpha$, for a
        set of values of the horizon radius $r_h$,  for hairy BHs of the $\mathrm{O}(3)$ model with $m=1$ and $\omega=0.9$.}
    \lbfig{alpha}
\end{figure}

Firstly, we do not find evidence for a maximal critical value of $\alpha$; $i.e.$, the solutions appear to exist for arbitrary large values of the
coupling. Secondly, both the scaled mass  and the Hawking temperature of the configurations remain almost
constant as $\alpha$ increases above a certain
critical value. In other words, in the limit of strong gravitational coupling, the ADM mass
of the solutions decreases as $M\sim\alpha^{-2}$. The scale invariance of the model is effectively restored
and the profile functions of the solutions become
completely independent on the strength of the gravitational interactions. Notably, this pattern is
also observed in the limit $\alpha\to \infty$ both for the topologically trivial pion clouds in the
Einstein-Skyrme model \cite{Ioannidou:2006nn}, and for rotating boson stars \cite{Kleihaus:2005me}.

For solitonic solutions there are, generally, two $\alpha$-branches of solutions, which bifurcate at some value of the gravitational coupling.
However, as the angular velocity becomes relatively high, $\omega \sim 0.95$, we observe just
a single branch of solitonic solutions.
This branch terminates at some small finite value of the coupling
$\alpha$, wherein the mass and the
angular momentum diverge like $\alpha^{3}$. Indeed, in the $\mathrm{O}(3)$ non-linear sigma model the
spinning solitons do not possess a non-trivial flat space limit.

For the hairy BHs we can also find two $\alpha$-branches of solutions, which, for small values
of $r_h$, are disconnected -  Fig.~\ref{alpha}. As $r_h$ increases, the branches merge at some minimal value
$\alpha_\mathrm{min}>0$; these solutions exist only as  $\alpha>\alpha_\mathrm{min}$. The value of  $\alpha_\mathrm{min}$
rapidly increases as $r_h$ increases.

Up to now we have always been discussing solutions with $m=1$. Let us briefly mention now solutions with higher windings $m>1$ - Fig.~\ref{m}.
\begin{figure}[hbt]
    \begin{center}
        \includegraphics[width=.48\textwidth, trim = 40 20 90 20, clip = true]{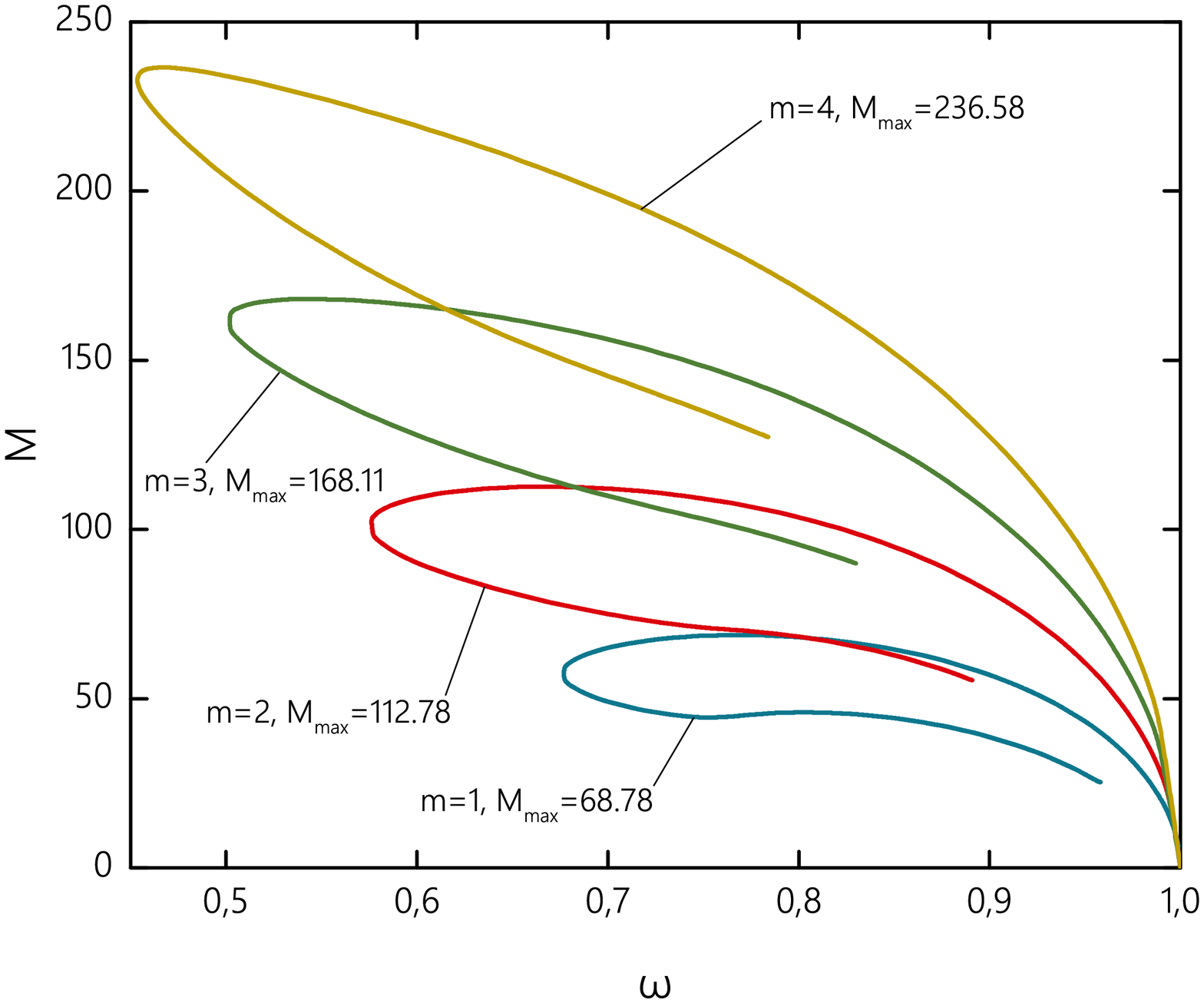}
        \includegraphics[width=.48\textwidth, trim = 40 20 90 20, clip = true]{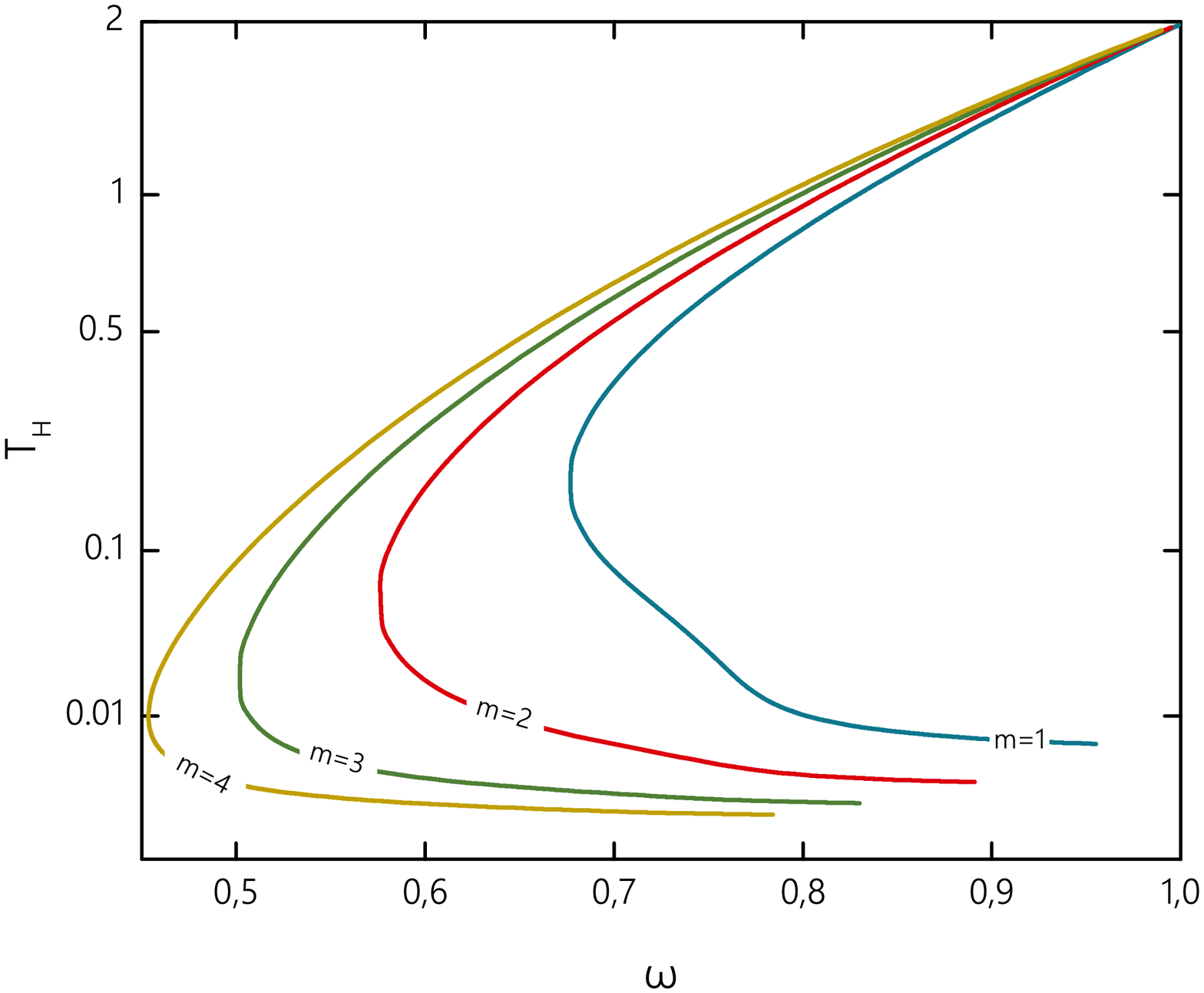}
        \includegraphics[width=.6\textwidth, trim = 40 20 90 20, clip = true]{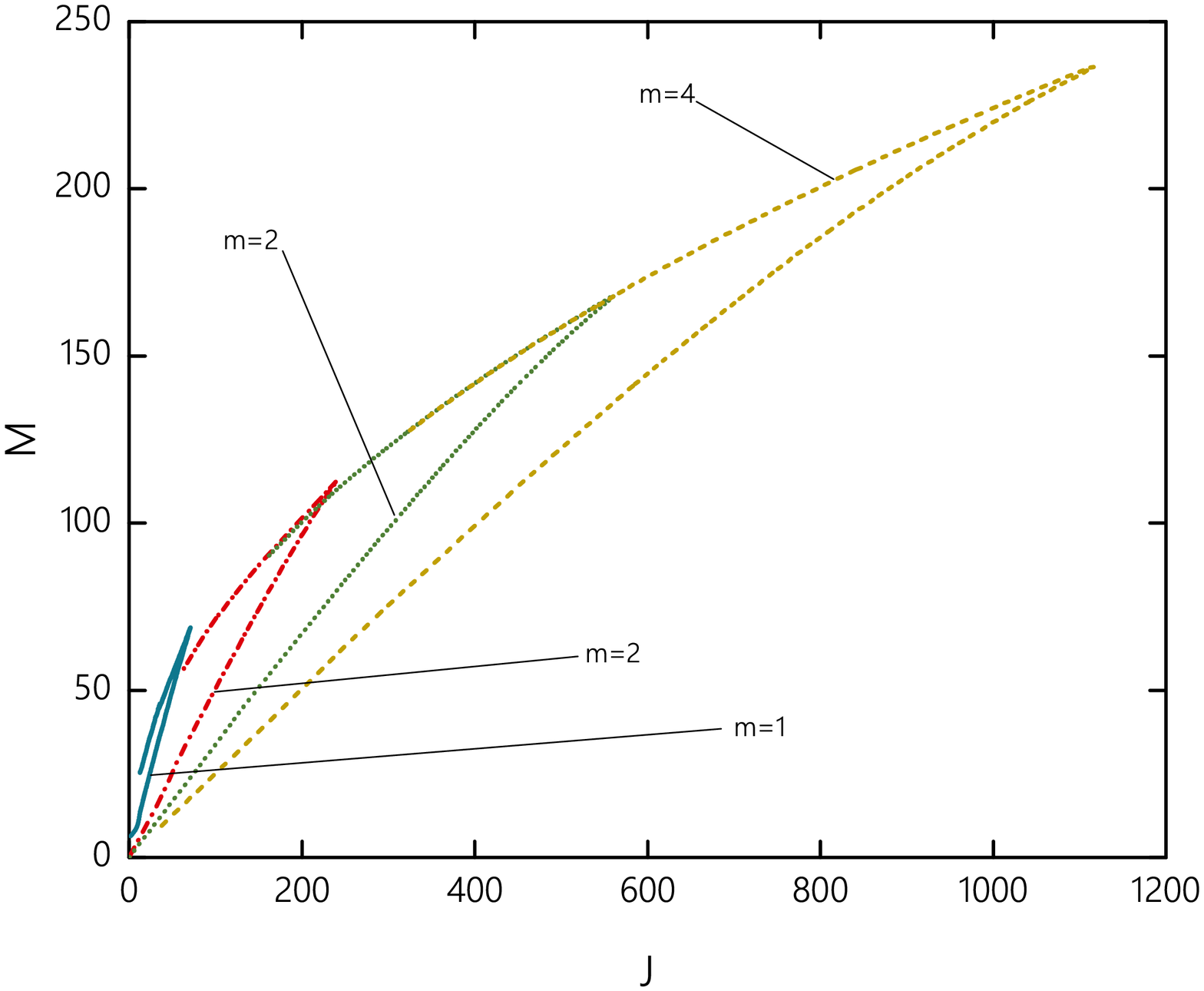}
    \end{center}
    \caption{\small
        ADM mass $M$ (upper left panel) and  Hawking temperature $T_H$ (upper right panel)
        as functions  of the angular frequency $\omega$.  Scaled ADM mass $vs.$ total angular momentum $J$ (bottom panel)
        for a set of values of the winding number $m$ for rotating hairy BHs
        with $r_h=0.01, \alpha=0.5$.}
    \lbfig{m}
\end{figure}
We have observed their behavior, generically, mimics the same qualitative pattern of the  $m=1$ configurations.
As basic trends, as the winding $m$ increases, both the ADM mass and the angular momentum increase, while the minimal allowed value of the angular frequency $\omega_{\mathrm{min}}$ decreases. Fig.~\ref{m} presents the variation of the ADM mass and Hawking temperature with $\omega$ and the mass-angular momentum relation for some selected hairy BH solutions with fixed $r_h=0.01$ and $m=1,2,3,4$.

Finally, let us briefly comment on the geometry of the horizon of the hairy BHs. In Fig.~\ref{ergo}, the spatial sections of the even horizon are represented as a round $S^2$. Geometrically, however, these surfaces are squashed 2-spheres, rather round ones. To quantify this deformation, we have evaluated the ratio of equatorial to polar 
circumferences\footnote{For the considered metric ansatz (\ref{metrans}), one defines 
$L_e\equiv 2\pi r_h \sqrt{F_2(r_h,\pi/2}$,
and
$L_p\equiv 2r_h\int_0^{\pi}d\theta \sqrt{F_1(r_h,\theta)}$.
}, 
$\epsilon = L_e/L_p$, which is exhibited in Fig.~\ref{LeLp}.
Along the first forward branch the value of $\epsilon$ is slightly smaller than 1, weakly decreasing as $\omega$ decreases. Hence, as mentioned above the BHs are Schwarzschild-like in the vicinity of the Newtonian limit.  The deformation becomes considerably stronger along the secondary branches, towards the Kerr limit.

\begin{figure}[hbt]
    \begin{center}
        \includegraphics[width=.48\textwidth, trim = 40 20 90 20, clip = true]{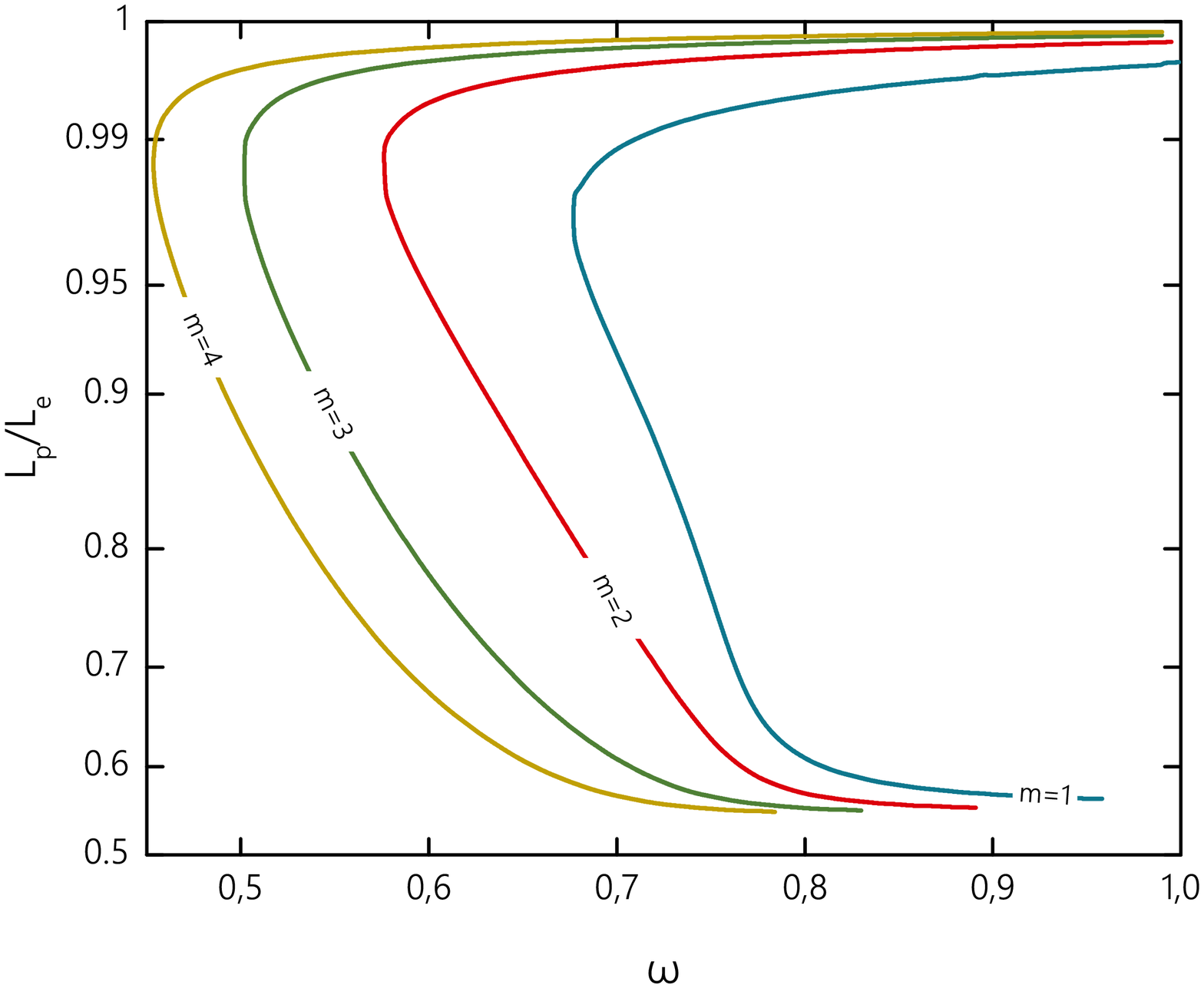}
        \includegraphics[width=.48\textwidth, trim = 40 20 90 20, clip = true]{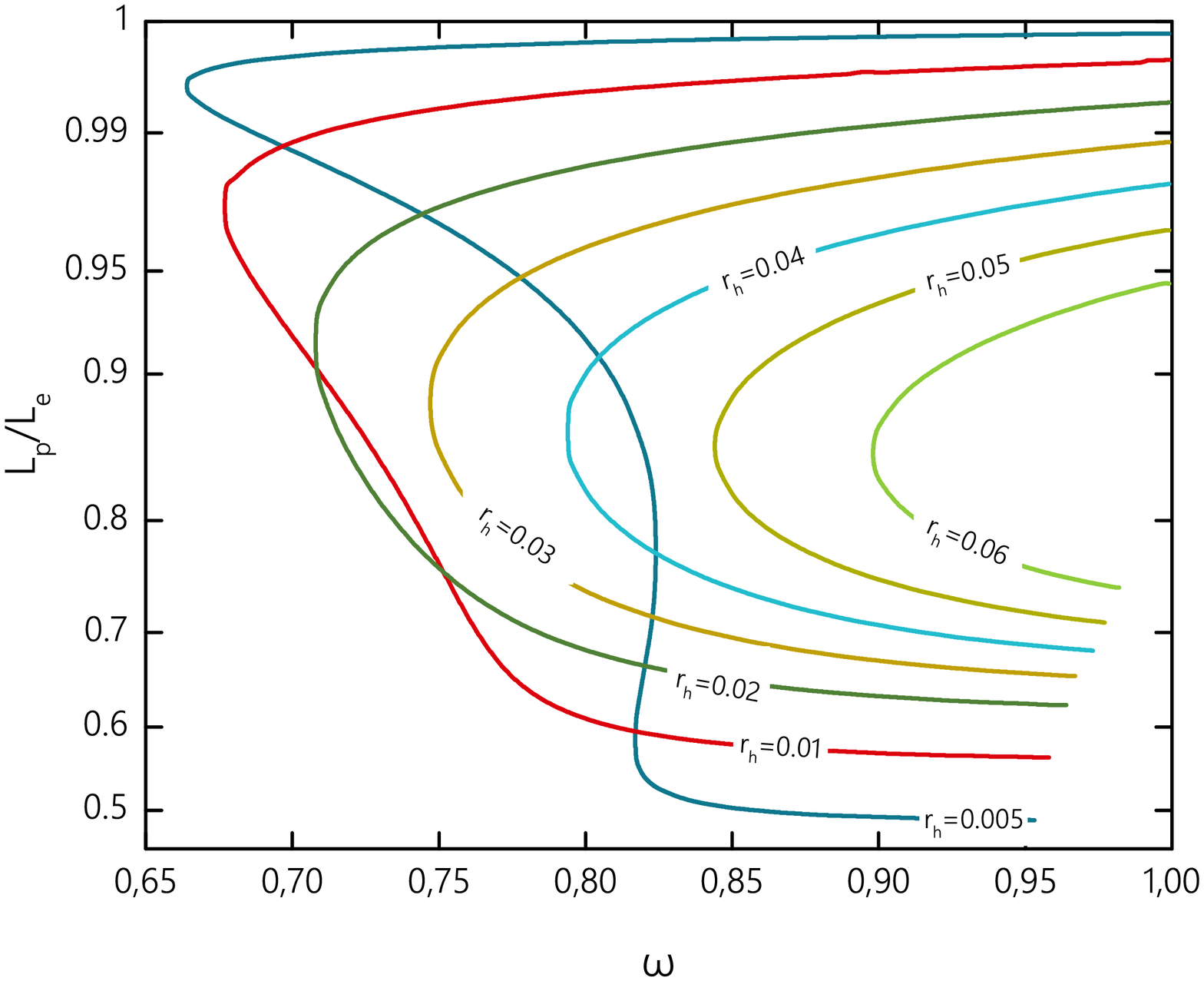}
    \end{center}
    \caption{\small The ratio of the horizon circumferences $L_e/L_p$ is shown as a function of: winding number $m$ (left panel), for fixed $r_h=0.01$;  angular frequency $\omega$ (right panel) for the forward branch solutions with $m=1$. Here all solutions have $\alpha=0.5$.
    }
    \lbfig{LeLp}
\end{figure}

\section{Conclusions}
\label{section3}

In this paper we have shown that the $\mathrm{O}(3)$ non-linear sigma-model, a simple and geometrically motivated field theory that does not admit solitonic solutions in flat spacetime, in the absence of higher order kinetic terms, can admit solitonic and hairy BH solutions when minimally coupled to gravity and when in the presence of symmetry non-inheritance. A close resemblance with the case of the massive, complex Klein-Gordon field theory has been observed. In the latter, no soliton like solutions exist in flat spacetime and, again, coupling to gravity can yield self-gravitating (both static and spinning) solitonic solutions - boson stars - and spinning hairy BHs. In fact, a parallelism on the structure of the domain of existence of solutions and physical properties between the $\mathrm{O}(3)$ sigma model and the Einstein-Klein-Gordon model has been clearly exhibited.

It is well known  that field theories, in order to possess solitonic type solutions, need non-linearities. This is a necessary but not sufficient condition as shown by the absence  of non-trivial solutions in the pure $\mathrm{O}(3)$ non-linear sigma-model.
The soliton enabling non-linearities can have different origins:
{\bf i)} self-interactions of the field, as illustrated by $Q$-balls in flat spacetime;
{\bf ii)} higher order kinetic terms, as illustrated by the Skyrme model;
{\bf iii)} coupling to Einstein's gravity, as illustrated for mini-boson stars in the massive-complex-Klein-Gordon model.
For the $\mathrm{O}(3)$ non-linear sigma model case, it is well known that in the presence of higher order kinetic terms it
gives rise to the topologically non-trivial Hopfions~\cite{Fadeev1,Faddeev:1976pg,Gladikowski:1996mb,Manton:2004tk,Radu:2008pp,Shnir:2018yzp} (point {\bf ii}); here we have shown that the coupling to gravity gives rise to new regular,
topologically trivial spherically symmetric and spinning solitons (point {\bf iii)}, and, via the synchronisation mechanism,
to BHs with synchronised hair.
As pointed out above, using a harmonic time dependence in field space and a sufficiently accommodating potential, it should also contain  solitonic solutions in flat spacetime (point {\bf i)}.
 It would be interesting to study such solutions. One observes that, as a general pattern, symmetry non-inheritance accompanies the existence of non-topological solitons.

From the viewpoint of the hairy BH solutions found in this work, it again shows the universality of
the synchronisation mechanism to yield hairy BHs in a large variety of field theory models
(see another example in~\cite{Herdeiro:2016tmi}), in different dimensions and asymptotics - see
$e.g.$~\cite{Dias:2011at,Brihaye:2014nba}. Thus, this mechanism can be extended to
non-linear sigma models when coupled to gravity.

Finally, we remark that the $\mathrm{O}(3)$ configurations reported in this work
are topologically trivial,
 being constructed for the simplest scalar fields ansatz (\ref{scalans}).
However, more general solutions,
which carry a nonzero  Hopf charge density,
should also exist
within  the same model (\ref{lag}),
by considering an extended $\mathrm{O}(3)$ ansatz with two essential functions. Further, a similarity of the model with
Einstein-Skyrme theory \cite{Kleihaus:2005me}
suggests that the pattern of the spinning self-gravitating solutions may be very involved, in
particular we can expect the topologically trivial clouds may be bounded by isospinning Hopfion forming additional branches
of solutions. We hope to address these problems in our future work.

\section*{Acknowledgements}
This work has been supported by the FCT (Portugal) IF programme, by the FCT grant PTDC/FIS-OUT/28407/2017,
by  CIDMA (FCT) strategic project UID/MAT/04106/2013, by CENTRA (FCT) strategic project UID/FIS/00099/2013 and
by  the  European  Union's  Horizon  2020  research  and  innovation  (RISE) programmes H2020-MSCA-RISE-2015
Grant No.~StronGrHEP-690904 and H2020-MSCA-RISE-2017 Grant No.~FunFiCO-777740. E.R. gratefully acknowledges the support of DIAS.
Ya.S. gratefully acknowledge support from the
Ministry of Education and Science of Russian Federation, project No 3.1386.2017 and from DAAD-Ostpartnerschaftsprogramm.
Ya.S. is grateful to Jutta Kunz and Burkhard Kleihaus for useful discussions. He would like to thank the
Kavli Institute for Theoretical Physics at University of California Santa Barbara for its kind
hospitality during the completion of this work. The authors would like to acknowledge
networking support by the COST Action CA16104.

\begin{small}

\end{small}

\end{document}